\documentclass[12pt]{article}

\PassOptionsToPackage{hyperfootnotes=false}{hyperref}
\usepackage{lmodern}
\usepackage{textcomp}
\usepackage{bakerchet}
\DeclareFontShape{T1}{cmr}{x}{n}{<->cmr10}{}
\renewcommand{\mytitlefont}{\fontfamily{cmr}\fontseries{x}\selectfont}
\usepackage{amssymb,amsmath,amsfonts,mathrsfs,braket,mathtools}
\usepackage{pgfplots}
\usepackage[utf8]{inputenc}
\usepackage{graphicx,longtable,afterpage,dsfont}
\usepackage[T1]{fontenc}
\usepackage{xcolor,hyperref}
\usepackage{xurl}

\usepackage{subcaption}
\usepackage{microtype}
\usepackage{booktabs}
\usepackage{multirow}
\usepackage{hyphenat}
\usepackage{mathtools}
\usepackage[capitalize]{cleveref}
\usepackage[normalem]{ulem}
\usepackage{tikz}
\usepackage{setspace}
\usepackage{tensor}
\usepackage{todonotes}
\allowdisplaybreaks

\newcommand{\OO}{\mathcal{O}}
\newcommand{\GG}{\mathcal{G}}
\newcommand{\el}{\ell}
\newcommand{\dd}{\mathrm{d}}
\newcommand{\bx}{\boldsymbol{x}}
\newcommand{\bk}{\boldsymbol{k}}

\newcommand{\MM}{\mathcal{M}}
\newcommand{\NN}{\mathcal{N}}

\def\KK{\mathcal{K}}
\def\JJ{\mathcal{J}}

\def\BB{{\cal B}}

\def\GG{{\cal G}}

\def\JJ{{\cal J}}
\def\KK{{\cal K}}

\def\MM{{\cal M}}
\def\NN{{\cal N}}
\def\OO{{\cal O}}

\def\QQ{{\cal Q}}
\def\RR{{\cal R}}
\def\SS{{\cal S}}

\def\WW{{\cal W}}
\def\XX{{\cal X}}

\def\IR{{\mathbb R}}

\def\d{{\partial}}
\def\pp{{p_n}}

\usetikzlibrary{shapes.geometric, arrows, arrows.meta, backgrounds, fit, positioning,calc}

\tikzstyle{startstop} = [ellipse, draw, fill=green!20, text width=6cm, minimum height=1cm, align=center]
\tikzstyle{process} = [rectangle, draw, fill=blue!10, text width=6cm, minimum height=1cm, align=center]
\tikzstyle{decision} = [diamond, draw, fill=yellow!20, text width=4cm, minimum height=1cm, align=center, aspect=2]
\tikzstyle{parallel} = [rectangle, draw, fill=cyan!10, text width=6cm, minimum height=1cm, align=center]

\pgfplotsset{compat=1.16}
\preprint{QMUL-PH-26-25 \hfill CCTP-2026-15  \\ ITCP-IPP 2026/15}

\title{\fontsize{20}{19}\selectfont
  Thermal Double-Twist Data in Holography
}

\author{
V.~Niarchos\;$^{a,\bigstar}$, 
C.~Papageorgakis\;$^{b,\blacklozenge}$ and 
A.~Stratoudakis\;$^{a,c,\spadesuit}$}

\affiliation{
$^a$ Department of Physics, ITCP \& CCTP,\\
University of Crete, 71003 Heraklion, Greece \\
$^b$ Centre for Theoretical Physics and Astronomy\\ 
School of Physical and Chemical Sciences\\ Queen Mary University of London, London E1 4NS, UK\\
$^c$ Fields and Strings Laboratory, Institute of Physics\\ \'Ecole Polytechnique F\'ed\'erale de Lausanne
(EPFL)\\ Route de la Sorge, CH-1015 Lausanne, Switzerland \vspace{0.3cm} $ $ \\
\vspace{0.3cm} $ $\\

\vspace{0.0cm}
{\tt \small
$^\bigstar$niarchos@physics.uoc.gr,
$^\blacklozenge$c.papageorgakis@qmul.ac.uk, 
$^\spadesuit$alexandros.stratoudakis@epfl.ch}
\vspace{0.5cm}
}
\abstract{\noindent
We explain how to extract thermal OPE coefficients of double-twist operators in scalar two-point functions at infinite spatial volume from suitably regularized integrals of thermal response functions in momentum space. As a specific application, we implement the proposed approach to four-dimensional holographic CFTs with an Einstein bulk action, where the finite-temperature state is captured by an AdS$_5$ black-brane geometry. In that case, the thermal response function can be computed on the gravitational side by solving numerically a radial ODE reduction of the Klein--Gordon equation. We obtain accurate values of double-twist data from this solution, completing previous holographic and bootstrap studies of thermal two-point functions. Some of the reported spin-resolved data that we compute are new.
}

\date{\today}

\begin{document}

\hypersetup{pageanchor=false}
\maketitle
\hypersetup{pageanchor=true}

\setcounter{tocdepth}{2}
\toc

\newpage

\section{Introduction and Summary}
\label{sec:intro}

We consider $d$-dimensional conformal field theories (CFTs) on $S^1_\beta \times \IR^{d-1}$. The $S^1_\beta$ is a thermal circle with inverse temperature $\beta$. We are interested in the structure of the Operator Product Expansion (OPE) of scalar two-point functions $\langle \OO(x) \OO(0)\rangle_\beta$ on this geometry for generic scalar operators $\OO$ with scaling dimension $\Delta_\OO$. When the OPE is inserted inside the two-point function, it allows us to re-express the latter as a series of terms that are proportional to the thermal one-point functions of primary operators. These thermal one-point functions encode useful information about the thermal properties of the CFT and are the main target of this paper. 

When the CFT admits a holographic limit at large central charge $c$, the OPE of a thermal scalar two-point function is dominated at leading order in $1/c$ by the contributions of the identity operator, double-twist operators $[\OO\OO]_{s,J}$ (schematically of the form $\OO \Box^s \d^{\mu_1}\cdots \d^{\mu_J} \OO$), the energy-momentum tensor $T^{\mu\nu}$ and multi-trace composites thereof, $[T^m]_J$, without derivatives \cite{Iliesiu:2018fao}. The holographic importance of the energy-momentum sector is immediately obvious: it maps directly to graviton physics in the dual AdS black-hole geometry. Nevertheless, a fully consistent Kubo--Martin--Schwinger (KMS)-invariant two-point function is impossible without the double-twist sector and its presence is crucial in subtle geometric effects, like the resolution of the bouncing singularity in $d\geq 3$ CFTs \cite{Fidkowski:2003nf,Festuccia:2005pi,Festuccia:2008zx,Ceplak:2024bja}.   

The thermal one-point functions of the energy-momentum sector can be computed cleanly from gravity \cite{Fitzpatrick:2019zqz} (see also \cite{Buric:2025fye} for an improved computational scheme).\footnote{We refer the reader to various follow-up works in Refs.~\cite{Karlsson:2019dbd,Karlsson:2022osn,Fitzpatrick:2020yjb,Huang:2023ikg,Huang:2024wbq,Esper:2023jeq,Li:2019tpf,Li:2019zba,Li:2020dqm}.} Crucially, these data can be extracted from the near-boundary geometry without the need for detailed information of the solution of the scalar Klein--Gordon equation in the bulk. By contrast, the extraction of the double-twist thermal one-point functions is much harder because these data are sensitive to the full profile of the scalar perturbation in the bulk, including its behavior near the horizon. Previous work, \cite{Parisini:2023nbd,Buric:2025fye}, attempted to determine these data by fitting the OPE expansion to the two-point function in real space obtained from the bulk by solving a PDE---a difficult computation limited in accuracy.\footnote{A holographic approach that computes the Fourier series representation of the two-point function at fixed spatial momentum was presented more recently in \cite{Arnaudo:2026thermal}.} An alternative non-gravitational thermal bootstrap approach based on the KMS condition and a Borel--Pad\'e resummation was employed in \cite{Buric:2025anb,Buric:2025fye}. That work focused on the kinematic regime of zero-spatial separation, where the OPE contributions to the two-point thermal correlator wash away any spin-dependent information.\footnote{Additional recent thermal-bootstrap work relevant to holographic theories includes  \cite{Barrat:2025nvu,Barrat:2025twb}.} Even at non-zero spatial separation, the leading thermal data $a_{1,0}$ and $a_{0,2}$ for the double-twist operators $[\OO\OO]_{1,0}$ and $[\OO\OO]_{0,2}$ respectively, contribute together through the combination $a_{1,0}+\frac{1}{2}(d-2)(d-1)a_{0,2}$ to the KMS condition and cannot be resolved \cite{Niarchos:2025cdg}. The KMS condition is also agnostic to the thermal OPE coefficient $a_{0,0}$ of the simple double-twist operator $[\OO\OO]_{0,0}$. 

The present paper aims to fill the current gap in the holographic computation of thermal double-twist data by providing a new computational route. The main input (the thermal response function) still comes from a numerical solution of the bulk Klein--Gordon equation, but we do not attempt to determine directly the two-point function in physical space (by solving a PDE) and read off the OPE coefficients as in previous approaches. Instead, we compute the thermal response function in momentum space by solving a radial ODE reduction of the bulk Klein--Gordon equation and the solution becomes the input to a formula that relates each double-twist OPE datum individually to a regulated momentum-space integral of the thermal response function. The basis of this formula is a point-splitting definition of the double-twist operators together with the OPE that allows us to understand how to regulate the divergences in the coincident-point limit.    

Our analysis is general and applies to holographic CFTs in spacetime dimensions $d\geq 3$. As a specific illustration, we focus on the case of a four-dimensional holographic CFT with an AdS$_5$ black-brane dual and consider the thermal two-point function of a scalar primary operator with generic scaling dimension $\Delta_\OO$. To avoid well-known subtleties that arise when $\Delta_\OO$ is an integer (see e.g.~\cite{Fitzpatrick:2019zqz}, or Appendix E of \cite{Niarchos:2025cdg}) we assume for simplicity that $\Delta_\OO$ is a non-integer number above the unitarity bound. For example, at $\Delta_\OO = \frac{3}{2}$, we determine the three aforementioned lightest double-twist coefficients to high accuracy at the values
\begin{equation*}
  a_{0,0}=1.113079(1)\;,\qquad a_{1,0}=0.489(1)\;,\qquad a_{0,2}=2.399(1)\;.
\end{equation*}
The lightest scalar coefficient $a_{0,0}$ is a considerable refinement of an earlier holographic value~\cite{Parisini:2023nbd}, while $a_{1,0}$ and $a_{0,2}$ are, to our knowledge, the first computations of these individual coefficients. The combination that appears in the KMS condition, $a_{1,0} + 3 a_{0,2}$, is $7.686(2)$, in excellent agreement with the zero-spatial separation bootstrap analysis of~\cite{Buric:2025anb,Buric:2025fye}. All three coefficients differ markedly from their free-field values, and the difference is a direct measure of the black brane physics carried by the thermal state.

Nothing in our construction is special to the lightest operators reported above. The same approach
expresses every double-twist coefficient as a regulated momentum-space
integral of the thermal response function. Reaching the heavier double-twists is
essentially a matter of numerical control. As the operator dimension grows, the corresponding thermal data  probe the thermal response at
larger momenta, demanding a correspondingly more accurate solution of the radial
Klein--Gordon ODE and a more delicate regularization of the coincident-point
limit. 

The rest of the paper is organized as follows. \cref{sec:setup} sets up the bulk problem and extracts the thermal response from the black-brane geometry. In \cref{sec:deformations} we motivate the main formula for the double-twist coefficients by analyzing the generating functional of correlation functions with double-trace sources. In \cref{sec:equivalence} we implement a point-splitting definition of the double-twist one-point functions and determine the necessary regularization of the resulting expressions using the thermal-block expansion. \cref{sec:regularization} develops the subtraction and regularization scheme further and presents concrete numerical results. Two appendices collect the technical material used in the main text.

\section{Thermal response function from the \texorpdfstring{$AdS_5$}{AdS5} black brane}
\label{sec:setup}

We are working in five-dimensional Einstein gravity with a negative cosmological constant. The relevant Euclidean background is the planar AdS$_5$ black brane solution with metric
\begin{equation}\label{eq:ads5bh}
  \dd s^2 = \frac{\el^2}{z^2}\Bigl(f(z)\,\dd\tau^2 + \dd\bx^2
  + \frac{\dd z^2}{f(z)}\Bigr)\;,
  \qquad f(z)=1-\frac{z^4}{z_h^4}\;,
  \qquad T=\frac{1}{\pi z_h}\;,
\end{equation}
where $z=0$ is the conformal boundary, $z=z_h$ the horizon, $\el$ the AdS radius,
and the Euclidean time is periodic, $\tau\sim\tau+\beta$ with $\beta=1/T=\pi z_h$.
The zero-temperature, pure AdS$_5$ geometry is recovered in the limit $z_h\to\infty$
($f\to1$). This theory is assumed to be the gravitational dual of a four-dimensional CFT on $S^1_\beta\times\mathbb{R}^3$ at large central charge $c\sim N^2$ and strong coupling. 

The bulk theory also contains a free scalar field $\phi$ of mass $m$, which is dual to a scalar primary operator $\OO$ on the boundary with scaling dimension $\Delta_\OO$. The relation between bulk mass and the boundary scaling dimension follows from the standard identification
\begin{equation}\label{eq:massdim}
  m^2\el^2=\Delta_\OO(\Delta_\OO-4)\;.
\end{equation} 
We are interested in the thermal two-point function of the boundary operator \begin{equation}\label{eq:gdef}
g(\tau,\bx)\coloneqq\langle\OO(\tau,\bx)\,\OO(0,\boldsymbol{0})\rangle_\beta\;.
\end{equation}
 
\subsection{Zero-temperature warm-up}
\label{sec:T0}

To begin with, we consider the zero-temperature case, which allows us to fix the notation and the corresponding operator normalization. At quadratic order, the bulk scalar equation of motion is the Klein--Gordon equation on the AdS$_5$ background. Separating the bulk radial coordinate $z$ from the boundary coordinates $x^\mu=(\tau,\bx)$, we set
$\phi(x,z)=e^{ip\cdot x}\psi_p(z)$, with Euclidean momentum $p$ such that $p^2\ge 0$. This reduces the Klein--Gordon equation to the radial ODE
\begin{equation}\label{eq:radialODE}
  \psi''_p - \frac{3}{z}\psi'_p - \Bigl(p^2+\frac{m^2\el^2}{z^2}\Bigr)\psi_p=0\;.
\end{equation}
The substitution $\psi_p=z^2\chi_p$ maps this equation further to the modified Bessel equation of order $\nu = |\Delta_\OO -2|$, and the requirement of regularity in the interior ($z\to\infty$) selects the solution
\begin{equation}\label{eq:physsoln}
  \psi_p(z)=\ell^{-3/2} (pz)^2\,K_\nu(pz)\;,
\end{equation}
with $K_\nu$ the modified Bessel function of the second kind. Expanding near the boundary exposes the two
characteristic powers,
\begin{equation}\label{eq:nearbd}
  \psi(z)=\ell^{-3/2}A(p)\,z^{4-\Delta_\OO}+\cdots + \ell^{-3/2} B(p)\,z^{\Delta_\OO}+\cdots\;,
\end{equation}
which identify the source $A(p)$ and the response $B(p)$ in the holographic dictionary. For simplicity, we have assumed $\Delta_\OO\notin\mathbb{N}$.\footnote{For integer $\Delta_\OO$ one must consider carefully the presence of logarithmic terms in the near-boundary expansion~\eqref{eq:nearbd}.} In standard quantization ($\Delta_\OO>2$) the source is $A$ and the response $B$, giving the zero-temperature response function $\GG_0=2\nu B/A$. The case where $\Delta_\OO<2$ corresponds to the alternative quantization, in which the roles of $A$ and $B$ are exchanged and the response function is
\begin{equation}\label{eq:G0}
  \GG_0(p)=\frac{2\nu A(p)}{B(p)}=\frac{2\nu \Gamma(\nu)}{\Gamma(-\nu)}\Bigl(\frac{p}{2}\Bigr)^{-2\nu}\;,
\end{equation}
which decays as $p\to\infty$. The subscript 0 emphasizes that this is a zero-temperature quantity. 

Holographic renormalization of the
on-shell action (see \cref{app:holoren}) and the holographic dictionary yield the standard CFT two-point function. Upon adopting unit normalization in position space, we arrive at 
\begin{equation}\label{eq:2ptpositionmain}
  \langle\OO(x)\,\OO(0)\rangle
  =\frac{ 1}{|x|^{2\Delta_\OO}} =   \NN_{\Delta_\OO}\int\frac{\dd^4p}{(2\pi)^4}\,e^{ip\cdot x}\,\GG_0(p)\;,
\end{equation}
where
\begin{equation}\label{eq:Ndef}
  \NN_{\Delta_\OO}=\frac{\pi^2}{2\nu (\Delta_\OO-1)(\Delta_\OO-2)}\;.
\end{equation}

\subsection{The finite-temperature radial problem}
\label{sec:radialT}
At finite temperature, the separation of variables
\begin{equation}\label{eq:separation}  \phi(\tau,\bx,z)=e^{i\omega_n\tau+i\bk\cdot\bx}\,\psi_{n,k}(z)\;,
  \qquad \omega_n=\frac{2\pi n}{\beta}=\frac{2n}{z_h}\quad (n\in\mathbb{Z}),\quad k\coloneqq|\bk|\;,
\end{equation}
takes into account the periodicity in $\tau$ with $\omega_n$ the bosonic Matsubara frequencies. Upon substitution, the Klein--Gordon equation yields the following radial ODE
\begin{equation}\label{eq:radialBH}
  f\,\psi_{n,k}'' + \Bigl(f'-\frac{3}{z}f\Bigr)\psi_{n,k}'
  + \Bigl(-\frac{\omega_n^2}{f}-k^2+\frac{4-\nu^2}{z^2}\Bigr)\psi_{n,k}=0\;,
\end{equation}
with $f$ the blackening factor of \cref{eq:ads5bh}. Near the boundary $z=0$, this equation has a regular singular point with indices $\rho=2\pm\nu=\{4-\Delta_\OO,\,\Delta_\OO\}$, reproducing the source/response structure of~\cref{eq:nearbd}. The
associated Frobenius solution is
\begin{equation}\label{eq:bdr_frob}
  \psi_{n,k}(z)=\ell^{-3/2} A(n,k)\,z^{2-\nu}\sum_{m\ge0}a_m z^m
        +\ell^{-3/2} B(n,k)\,z^{2+\nu}\sum_{m\ge0}b_m z^m,\qquad a_0=b_0=1\;,
\end{equation}
with the relevant recursion relations between the coefficients collected in Appendix~\ref{app:frobenius}. It is straightforward to check that the odd-$m$ terms vanish and that the leading thermal corrections first appear at $m=4$.

To analyze the behavior near the horizon $z=z_h$, it is convenient to change variables to the dimensionless coordinate $u=z/z_h$ and rewrite the radial \cref{eq:radialBH} as
\begin{equation}\label{eq:radialBH_u}
  f\,\ddot\psi_{n,k} + \Bigl(\dot f-\frac{3}{u}f\Bigr)\dot\psi_{n,k}
  + \Bigl(-\frac{4n^2}{f}-(kz_h)^2+\frac{4-\nu^2}{u^2}\Bigr)\psi_{n,k}=0\;,
\end{equation}
where the dots denote $u$-derivatives. At the horizon, $u=1$, the indices are $\rho=\pm|n|/2$, and regularity selects
\begin{equation}\label{eq:hor_frob}
  \psi_{n,k}(u)=\ell^{-3/2}(1-u)^{|n|/2}\sum_{m\ge0}h_m(1-u)^m\;.
\end{equation}
The explicit form of the recursion relations between the $h_m$ coefficients is also relegated to Appendix~\ref{app:frobenius}. Horizon regularity fixes the bulk solution uniquely up to an overall normalization.

Connecting this horizon-regular solution to the boundary fixes the Frobenius
coefficients $A(n,k)$ and $B(n,k)$ in~\eqref{eq:bdr_frob}. Their overall
normalization cancels in the ratio, which defines the thermal response function (in standard quantization)
\begin{equation}\label{eq:GTdefeq}
  \GG_T(n,k)=\frac{2\nu B(n,k)}{A(n,k)}\;,
\end{equation}
the finite-temperature generalization of $\GG_0$. 

Holographic renormalization proceeds exactly as in the zero-temperature case (see \cref{app:holoren}) and gives the non-local renormalized action
\begin{equation}\label{eq:SrenBH}
  S_{\rm ren}^{T}=-\frac{1}{2\beta}\sum_{n}\int\frac{\dd^3k}{(2\pi)^3}\,
  A_n(k)\,\GG_T(n,k)\,A_{-n}(-k)\;,
\end{equation}
from which the thermal momentum-space two-point function follows,
$\langle\OO\OO\rangle\propto \GG_T$. 
We emphasize that \cref{eq:SrenBH} is written for standard quantization. For the alternative quantization the Legendre transform exchanges source and response, as already incorporated in the response function~\eqref{eq:G0}. With the operator normalization fixed at zero temperature, \cref{eq:SrenBH} gives the momentum-space two-point function
\begin{equation}\label{eq:2ptBH}
  \langle\OO(\omega_{n'},\bk')\,\OO(\omega_n,\bk)\rangle
  =(2\pi)^3\,\beta\,\NN_{\Delta_\OO}\,\GG_T(n,k)\,\delta_{n+n',0}\,\delta^{(3)}(\bk+\bk')\;.
\end{equation}
In position space, this becomes 
\begin{equation}\label{eq:gGT}
g(\tau,\bx)=\NN_{\Delta_\OO}\,\frac1\beta\sum_{n\in\mathbb{Z}}\int_{\mathbb{R}^3}\frac{\dd^3k}{(2\pi)^3}\,e^{ip_n\cdot x}\,\GG_T(n,k)\;,\qquad p_n=(\omega_n,\bk)\;.
\end{equation}
This is the finite-temperature analog of~\eqref{eq:2ptpositionmain}.

\subsection{Numerical evaluation of the thermal response}
\label{sec:GTnumeric}
The thermal response function \cref{eq:GTdefeq} will be a key ingredient in the upcoming sections, when we compute data associated with the thermal two-point function \cref{eq:gGT}. Unlike the $T=0$ case, ~\eqref{eq:radialBH_u} has no closed-form solution and we will extract the source and response coefficients $A,B$ numerically through the following series of steps:
\begin{enumerate}
  \item Initialize $\psi_{n,k}$ and $\dot\psi_{n,k}$ close to the horizon, $u=1-\epsilon_h$,
        using the near-horizon Frobenius series~\eqref{eq:hor_frob}.
  \item Integrate the radial equation numerically in its first-order form. More precisely, we evolve the vector
    \[
    Y(u)=
    \begin{pmatrix}
    \psi_{n,k}(u)\\
    \dot\psi_{n,k}(u)
    \end{pmatrix}
    \]
    from
    \begin{equation}
        u=1-\epsilon_h
    \quad \text{down to} \quad
    u=\epsilon_{\mathrm{bdr}}, \qquad \epsilon_h,\epsilon_{\rm bdr} \ll 1
    \end{equation}
    using SciPy's \texttt{solve\_ivp}, which defaults to double precision (float64).
    In this way, the bulk solution is fixed uniquely by the regularity condition at the horizon, up to an overall normalization.
  \item Near the boundary, the solution has the asymptotic form
  \begin{equation}\label{eq:nearbd_full}
    \psi_{n,k}(u) \sim A\,u^{4-\Delta_\OO}\Bigl(1 + a_2\,u^2 + \cdots\Bigr)
    + B\,u^{\Delta_\OO}\Bigl(1 + b_2\,u^2 + \cdots\Bigr)\;,
  \end{equation}
  where $a_m,b_m$ are the boundary Frobenius coefficients~\eqref{eq:bdr_frob}. The coefficients $A$ and $B$ are determined by matching in a small overlap region near the boundary, where both the numerical solution and the truncated Frobenius expansions are accurate. In practice this is carried out by a least-squares fit using both $\psi_{n,k}(u)$ and $\dot\psi_{n,k}(u)$ over several points in the fitting window, which is more stable than single-point matching.\footnote{As the two falloffs $u^{4-\Delta_\OO}$ and $u^{\Delta_\OO}$ span many orders of magnitude across the window, the columns of the fit matrix are rescaled to unit norm before the solve, keeping the linear system well-conditioned.}
\end{enumerate}

\begin{figure}[t!]
    \centering
    \begin{subfigure}[t]{0.49\textwidth}
        \centering
        \includegraphics[width=\textwidth]{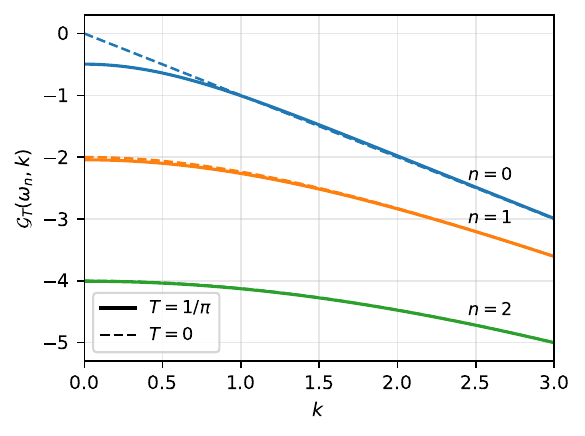}
        \caption{Standard response $\GG_T$ ($\Delta_\OO=\tfrac52$).}
    \end{subfigure}
    \hfill
    \begin{subfigure}[t]{0.49\textwidth}
        \centering
        \includegraphics[width=\textwidth]{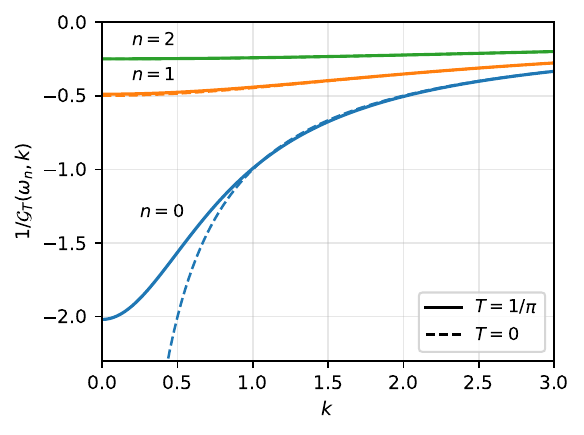}
        \caption{Alternative response $1/\GG_T$ ($\Delta_\OO=\tfrac32$).}
    \end{subfigure}
    \caption{The thermal response function in standard quantization (plot (a)) and its alternative-quantization counterpart (plot (b)) for $\nu=\tfrac12$, as functions of the Matsubara mode $n$ and the spatial momentum $k$.}
    \label{fig:GT}
\end{figure}

\paragraph{Example.}
\cref{tab:GT_ratio} reports the ratio $r(n,k)=\GG_T(n,k)/\GG_0(p)$, with
$p=\sqrt{4n^2+k^2}$, for $\nu=|\Delta_\OO-2|=\tfrac12$ ($\Delta_\OO=\tfrac32$ or $\tfrac52$, $m^2\el^2=-\frac{15}{4}$) in units $z_h=1$ (so that $T=1/\pi$). See also \cref{fig:GT}. The thermal response function approaches the zero-temperature result, $\GG_T\to\GG_0$, rapidly as $n$ or $k$ grows. The deviations are at the $0.1\%$ level already by $k=5$ and below $0.01\%$ by $k=10$. This will be a key feature underlying every subtraction used in \cref{sec:regularization}.

\begin{table}[t]
\centering
\begin{tabular}{c|cccc}
\hline
$n \backslash k$ & $1$ & $2$ & $5$ & $10$ \\
\hline
$0$ & $1.0064$ & $0.9901$ & $0.9997$ & $1.0000$ \\
$1$ & $1.0125$ & $1.0031$ & $0.9999$ & $1.0000$ \\
$2$ & $1.0017$ & $1.0010$ & $1.0001$ & $1.0000$ \\
$3$ & $1.0004$ & $1.0003$ & $1.0001$ & $1.0000$ \\
$4$ & $1.0001$ & $1.0001$ & $1.0000$ & $1.0000$ \\
\hline
\end{tabular}
\caption{Ratio $r(n,k)=\GG_T(n,k)/\GG_0(p)$ for $\nu=\tfrac12$, with
$p=\sqrt{4n^2+k^2}$, in units $z_h=1$. The horizon data are initialized with $10$
terms and the least-squares matching is performed in the window
$u\in[10^{-3},10^{-2}]$ using boundary Frobenius expansions up to
$\mathcal{O}(u^{20})$.}
\label{tab:GT_ratio}
\end{table}

\section{Double-twist data from double-trace deformations}
\label{sec:deformations}

In this section we develop a preliminary master formula that expresses the thermal one-point coefficients of the double-twist operators $[\OO\OO]_{s,J}$ in terms of the thermal response function $\GG_T$. The derivation of this formula is field-theoretic and goes through the study of the generating functional of double-trace deformations. This discussion will in turn serve as preparation for a more complete analysis, with properly regularized integrals, based on the OPE expansion in \cref{sec:equivalence}.  

The thermal two-point function $g(\tau,\bx)$ of Eq.~\eqref{eq:gdef} admits a convergent short-distance expansion in thermal blocks~\cite{Iliesiu:2018fao}, with radius of convergence $\beta$,
\begin{equation}\label{eq:thermal_OPE}
  g(\tau,\bx)=\sum_{\Delta,J}\frac{a_{\Delta,J}}{\beta^{\Delta}}\,C_J^{(\hat\nu)}(\eta)\,
  |x|^{-2\Delta_\OO+\Delta}\;,\quad |x|\coloneqq\sqrt{\tau^2+\bx^2}\;,\quad \eta\coloneqq\frac{\tau}{|x|}\;.
\end{equation}
The sum runs over primary operators of dimension $\Delta$ and even spins $J$ in the $\OO\times\OO$ OPE, while $C_J^{(\hat\nu)}(\eta)$ denote the Gegenbauer polynomials with $\hat\nu=(d-2)/2=1$ for $d=4$. The product $\beta^{-\Delta}C_J^{(\hat\nu)}(\eta)|x|^{-2\Delta_\OO+\Delta}$ is the position-space thermal block, while the coefficients $a_{\Delta,J}$ are the so-called thermal OPE data. For an exchanged operator $\OO'$ of scaling dimension $\Delta$ and spin $J$ the coefficients $a_{\Delta,J}$ combine the zero-temperature OPE coefficient $f_{\OO\OO\OO'}$, the thermal one-point coefficient
$b_{\OO'}$, and the two-point normalization $c_{\OO'}$ as follows~\cite{Iliesiu:2018fao}
\begin{equation}\label{eq:a_b_relation}
  a_{\Delta,J}=\frac{f_{\OO\OO\OO'}\,b_{\OO'}}{c_{\OO'}}\,\frac{J!}{2^{J}(\hat\nu)_J}\;,
\end{equation}
with $(\hat\nu)_J$ the Pochhammer symbol.

For a holographic CFT at large $N$ the spectrum splits into: i) the identity, ii) the double-twist tower
$[\OO\OO]_{s,J}\sim\OO\,\Box^{s}\mathscr{D}_J\OO$, with $\mathscr{D}_J^{\mu_1\cdots\mu_J}=\partial^{\mu_1}\cdots\partial^{\mu_J}-(\text{traces})$ being the
traceless-symmetric rank-$J$ derivative, and scaling dimension
\begin{equation}\label{eq:dt_dim}
  \Delta_{s,J}=2\Delta_\OO+2s+J\;,\qquad s,J\in\mathbb{Z}_{\ge0}\ (J\ \text{even})\;,
\end{equation}
and iii) the multi-stress-tensor tower $[T^m]_J$ (without derivatives)~\cite{Fitzpatrick:2019zqz,Niarchos:2025cdg}, with scaling dimension $\Delta=4m$ and even spin $0\le J\le 2m$. The OPE decomposition of the thermal two-point function then reduces to\footnote{As stated earlier, this holds when the external operator dimension is non-integer. Otherwise the decomposition develops logarithmic contributions.}
\begin{equation}\label{eq:OPE4d}
  g(\tau,\boldsymbol{x})=|x|^{-2\Delta_\OO}
  +\sum_{s,J}\frac{a_{s,J}}{\beta^{2\Delta_\OO+2s+J}}\,C_J^{(1)}(\eta)\,|x|^{2s+J}
  +\sum_{m,J}\frac{a^{(m)}_{J}}{\beta^{4m}}\,C_J^{(1)}(\eta)\,|x|^{-2\Delta_\OO+4m}\;.
\end{equation}
The lowest-twist multi-stress-tensor data of the second tower have been obtained from holography
in~\cite{Fitzpatrick:2019zqz}. 

In this paper we focus on the computation of the double-twist contributions and the corresponding coefficients $a_{s,J}$, which have been the hardest to obtain so far.

\subsection{Multi-trace deformations}
\label{sec:multitrace}

Consider the generating functional $W[\JJ]$ of connected single-trace correlation functions in a large-$N$ (boundary) CFT defined formally as the path integral over some collection of fundamental fields denoted by $\varphi$, with action $S[\varphi]$
\begin{equation}\label{eq:Z0}
  Z[\JJ]=e^{-W[\JJ]}=\int\mathcal{D}\varphi\,
  e^{-S[\varphi]-\int \dd^4x\, \JJ(x)\,\OO(x)}\;.
\end{equation}
$\OO$ is an arbitrary single-trace operator of our choice, $\JJ(x)$ the corresponding source and we assume the path integral is performed in the thermal state. We can extend this standard setup to include generic multi-trace operators $F(\OO)$ (where $F$ is, e.g., some monomial) by introducing additional sources $\lambda(x)$ for the multi-trace operators as\footnote{We assume a suitable implicit large-$N$ scaling of the multi-trace source that scales all terms in the action as $N^2$. In this sense, a single-trace source $\JJ(x)$ scales like $N$, a double-trace like $N^0$, and so on and so forth.} 
\begin{equation}\label{eq:Zlambda}
  Z[\JJ,\lambda]=e^{-\widetilde W[\JJ,\lambda]}
  =\int\mathcal{D}\varphi\,e^{-S[\varphi]-\int \dd^4x\,(\JJ(x)\OO(x)+\lambda(x) F(\OO)(x))}\;.
\end{equation}
We introduced the notation $\widetilde W$ to distinguish the case with multi-trace sources from the one involving only single-trace sources. A functional derivative with respect to $\lambda(x)$ and setting $\lambda = 0$ gives the thermal one-point function of the multi-trace operator $F(\OO)$ in the presence of the single-trace source $\JJ$,
\begin{equation}\label{eq:dWdlambda}
  \frac{\delta \widetilde W}{\delta\lambda(x)}\Big|_{\lambda\to 0}
  =\langle F(\OO)(x)\rangle_{\beta,\JJ}\;.
\end{equation}
Setting also $\JJ=0$ gives the thermal one-point function of the multi-trace operator in the original undeformed theory.

In this paper, we are interested in the double-twist operators $[\OO\OO]_{s,J}\sim \OO\,\Box^{s}\mathscr{D}_J\OO$. For that reason, in what follows we focus on generating functionals with double-trace deformations. As a warm-up, we first consider the scalar case of the simplest double-trace operator $F(\OO)=\tfrac12\OO^2$ and then move on to the general case of the double-twist operators $F(\OO)=\tfrac12\OO\,\Box^{s}\mathscr{D}_J\OO$.

\subsection{The scalar double-trace} 
\label{sec:O2}

For $F(\OO)=\tfrac12\OO^2$ we can linearize the double-trace deformation with the standard introduction of a Hubbard--Stratonovich field $\sigma$,\footnote{Here the Gaussian measure is normalized so that $\int\mathcal{D}\sigma\,e^{-\int \dd^4 x\,\frac{\sigma^2}{2\lambda}}=1$.}
\begin{equation}\label{eq:HS}
  e^{-\frac12\int \dd^4x\,\lambda(x)\OO^2(x)}=
  \int\mathcal{D}\sigma\,
  e^{-\int \dd^4 x\,(\frac{\sigma^2}{2\lambda}-i\sigma\OO)}\;.
\end{equation}
This manipulation allows us to recast the original path-integral of the full generating functional in terms of a shifted single-trace source and no double-trace deformation
\begin{equation}\label{eq:Seff}
  e^{-\widetilde W[\JJ,\lambda]}=\int\!\mathcal{D}\sigma\,e^{-S_{\rm eff}[\sigma,\JJ,\lambda]}\;,
  \qquad S_{\rm eff}[\sigma,\JJ,\lambda]=W[\JJ-i\sigma]+\int\dd^4x\,\frac{\sigma^2}{2\lambda}\;.
\end{equation}
Notice that $S_{\rm eff}$ involves the single-trace generating functional $W$ evaluated at the shifted source $\JJ-i\sigma$. The saddle-point equation $\sigma_*=i\lambda\langle\OO\rangle_{\widehat \JJ}$ involves in a non-trivial manner the shifted source
$\widehat \JJ \coloneqq \JJ-i\sigma_*=\JJ+\lambda\langle\OO\rangle_{\widehat \JJ}$. 
Expanding the path-integral over $\sigma$ in a power series around the saddle-point configuration $\sigma_*$ allows us to organize the functional $\widetilde W[\JJ,\lambda]$ in a $1/N$ expansion, which is convenient if we are working in the large-$N$ (holographic) limit. In this sense, expanding $S_{\rm eff}$ up to quadratic order in $\xi=\sigma-\sigma_*$ and integrating $\xi$ out gives the one-loop generating functional
\begin{equation}\label{eq:Wlambda_O2}
  \widetilde W[\JJ,\lambda]=W[\widehat \JJ]-\tfrac12 \int\dd^4y\,\lambda(y)\,\langle\OO(y)\rangle_{\widehat \JJ}^2
  +\tfrac12\,\mathrm{Tr}\log\!\bigl(1+\lambda\mathscr{G}_{\widehat \JJ}\bigr)+\mathcal{O}(N^{-2})\;,
\end{equation}
where $\mathscr{G}_{\widehat \JJ}$ is the undeformed connected two-point function at source $\widehat \JJ$. The $\mathrm{Tr}\log$ is a functional trace over $S^1_\beta\times\mathbb{R}^3$, in which $\mathscr{G}_{\widehat \JJ}(x-y)$ acts by convolution while $\lambda(x)$ acts by multiplication. For a spacetime-dependent $\lambda$ the operator $1+\lambda\mathscr{G}_{\widehat \JJ}$ is not diagonal in momentum space. We therefore keep the $\mathrm{Tr}\log$ as a functional trace and differentiate it directly. 

The thermal one-point function of the deformation follows from the functional derivative of \eqref{eq:Wlambda_O2} with respect to $\lambda(x)$ at $\JJ,\lambda\to0$. According to \eqref{eq:dWdlambda} with $F(\OO)=\tfrac12\OO^2$,
\begin{equation}\label{eq:dWO2}
  \frac{\delta\widetilde W}{\delta\lambda(x)}\bigg|_{\JJ,\lambda\to0}=\tfrac12\langle\OO^2(x)\rangle_\beta^{\rm bare}\;,
\end{equation}
where the superscript `bare' is a reminder that the resulting expression may have divergences that need to be regularized. Note also that the shifted source itself depends on $\lambda$ through the self-consistency relation $\widehat\JJ(y)=\JJ(y)+\lambda(y)\langle\OO(y)\rangle_{\widehat\JJ}$. Differentiating it with respect to $\lambda(x)$, and using $\delta\langle\OO(y)\rangle_{\widehat\JJ}/\delta\widehat\JJ(z)=-\mathscr{G}_{\widehat\JJ}(y,z)$, we obtain
\begin{equation}\label{eq:dJhat}
  \frac{\delta\widehat\JJ(y)}{\delta\lambda(x)}
  =\delta^{(4)}(x-y)\,\langle\OO(y)\rangle_{\widehat\JJ}
  -\lambda(y)\!\int\!\dd^4z\,\mathscr{G}_{\widehat\JJ}(y,z)\,\frac{\delta\widehat\JJ(z)}{\delta\lambda(x)}
  \;\xrightarrow{\;\lambda\to0\;}\;
  \delta^{(4)}(x-y)\,\langle\OO\rangle_\beta\;.
\end{equation}
With this input, differentiating the three terms of \eqref{eq:Wlambda_O2} with respect to $\lambda(x)$, using $\delta W/\delta\widehat\JJ(y)=\langle\OO(y)\rangle_{\widehat\JJ}$, gives
\begin{align}\label{eq:dWfull}
  \frac{\delta\widetilde W}{\delta\lambda(x)}
  =&\int\dd^4y\,\langle\OO(y)\rangle_{\widehat\JJ}\,\frac{\delta\widehat\JJ(y)}{\delta\lambda(x)}
  \;-\;\tfrac12\langle\OO(x)\rangle_{\widehat\JJ}^2
  \;-\;\int\dd^4y\,\lambda(y)\,\langle\OO(y)\rangle_{\widehat\JJ}\,
       \frac{\delta\langle\OO(y)\rangle_{\widehat\JJ}}{\delta\lambda(x)}
  \nonumber\\[0.4ex]
  &+\;\tfrac12\,\mathrm{Tr}\!\left[\frac{1}{1+\lambda\mathscr{G}_{\widehat\JJ}}\,
       \frac{\delta(\lambda\mathscr{G}_{\widehat\JJ})}{\delta\lambda(x)}\right]\;.
\end{align}
The numerator of the last term splits into an explicit and an implicit piece,
\begin{equation}\label{eq:dlamG}
  \frac{\delta(\lambda(u)\mathscr{G}_{\widehat\JJ}(u,v))}{\delta\lambda(x)}
  =\underbrace{\delta^{(4)}(x-u)\,\mathscr{G}_{\widehat\JJ}(u,v)}_{\text{explicit}}
  +\underbrace{\lambda(u)\!\int\!\dd^4z\,\frac{\delta\mathscr{G}_{\widehat\JJ}(u,v)}{\delta\widehat\JJ(z)}\,
       \frac{\delta\widehat\JJ(z)}{\delta\lambda(x)}}_{\text{implicit}}\;,
\end{equation}
the implicit piece arising because $\mathscr{G}_{\widehat\JJ}$ is evaluated at the $\lambda$-dependent shifted source.
At $\JJ,\lambda\to0$, $1/(1+\lambda\mathscr{G}_{\widehat\JJ})\to1$ and every remaining term that carries an explicit factor of $\lambda$ drops out, assuming proper analytic behavior for small deformations. As a result, the third term of \eqref{eq:dWfull} vanishes. For the surviving trace-log piece the trace of the explicit kernel in \eqref{eq:dlamG} reduces to the two-point function
\begin{equation}\label{eq:trlogderiv}
  \tfrac12\,\mathrm{Tr}\!\left[\frac{1}{1+\lambda\mathscr{G}_{\widehat\JJ}}\,
       \frac{\delta(\lambda\mathscr{G}_{\widehat\JJ})}{\delta\lambda(x)}\right]\!\Bigg|_{\JJ,\lambda\to0}
  =\tfrac12\,\mathscr{G}_{\widehat\JJ}(x,x)\big|_{\widehat\JJ\to0}
  =\tfrac12\int_p\mathscr{G}_{\widehat\JJ}(p)\big|_{\widehat\JJ\to0}\;,
\end{equation}
at the coincident point. Here we defined $\int_p\equiv\tfrac1\beta\sum_n\int\frac{\dd^3k}{(2\pi)^3}$ and the last equality is the coincident-point limit expressed in momentum space, $\mathscr{G}_{\widehat\JJ}(x,x)=\int_p\mathscr{G}_{\widehat\JJ}(p)$. Therefore, using \eqref{eq:dJhat} for the first term in \eqref{eq:dWfull}, the four contributions on the RHS of \eqref{eq:dWfull} collapse respectively to $\langle\OO\rangle_\beta^2$, $-\tfrac12\langle\OO\rangle_\beta^2$, $0$, and \eqref{eq:trlogderiv}. Collecting all the terms, we obtain
\begin{equation}\label{eq:O2half}
  \tfrac12\langle\OO^2(x)\rangle_\beta^{\rm bare}
  =\frac{\delta\widetilde W}{\delta\lambda(x)}\bigg|_{\JJ,\lambda\to0}
  =\tfrac12\langle\OO\rangle_\beta^2+\tfrac12\int_p\mathscr{G}_{\widehat\JJ}(p)\big|_{\widehat\JJ\to0}\;,
\end{equation}
which is position-independent, as expected for a thermal one-point function in a translation-invariant vacuum. Assuming we are in a vacuum where the single-trace one-point function vanishes, $\langle\OO\rangle_\beta=0$, the one-loop term carries the entire answer. 
The superscript `bare' on the LHS still highlights the fact that this is a formal divergent expression that requires regularization. Ignoring this aspect for a moment, we can insert the expression $\mathscr{G}_{\widehat\JJ}(n,k)|_{\widehat\JJ\to0}=\NN_{\Delta_\OO}\GG_T(n,k)$ from~\eqref{eq:2ptBH} to rewrite \eqref{eq:O2half} as
\begin{equation}\label{eq:O2bare}
  \langle\OO^2\rangle_\beta^{\rm bare}  =\NN_{\Delta_\OO}\,\frac1\beta\sum_{n\in\mathbb{Z}}\int_{\mathbb{R}^3}\frac{\dd^3k}{(2\pi)^3}\,\GG_T(n,k)\;.
\end{equation}

As we have just mentioned, this expression is divergent. In position space, it is the coincident-point value $g(0,\boldsymbol{0})$ of the thermal two-point function~\eqref{eq:gGT}, which according to the OPE~\eqref{eq:OPE4d} diverges at the coincident-point limit $|x|\to 0$ as $|x|^{-2\Delta_\OO}$. The divergence comes from the identity contribution to the OPE and the corresponding regularization relies on the proper point-splitting definition of the double-trace operator $\OO^2(x)$ with a suitable subtraction. This can be achieved by subtracting the zero-temperature two-point correlator 
\begin{equation}\label{eq:a00_def}
 \langle\OO^2\rangle_\beta^{\rm ren} =\NN_{\Delta_\OO}\!\left(\frac1\beta\sum_{n\in\mathbb{Z}}\int_{\mathbb{R}^3}\frac{\dd^3k}{(2\pi)^3}\GG_T(n,k)
  -\int_{\mathbb{R}^4}\frac{\dd^4p}{(2\pi)^4}\GG_0(p)\right)~. 
\end{equation}

This one-point function is closely related to the thermal OPE coefficient $a_{0,0}$, as we explain in more detail in \cref{sec:equivalence}. The main lesson of the above exercise is that the thermal one-point function of a double-trace operator can be extracted from the thermal response function with a proper point-splitting procedure.

\subsection{General double-trace deformations}
\label{sec:gendeform}

A similar analysis can be applied to the higher double-twist operators, which remain quadratic in $\OO$. For brevity, we omit obvious intermediate steps and highlight the key new aspects of the argument. In the general case, the local double-twist deformations 
\eqref{eq:Zlambda} 
involve a spacetime-dependent tensorial source
\begin{equation}\label{eq:gendeform}
  \frac12\int \dd^4x\,\dd^4y\,
  \lambda^{\mu_1\cdots\mu_J}(x)\,\OO(x)\,\KK_{\mu_1\cdots\mu_J}(x,y)\,\OO(y)\;,\qquad
  \KK(x,y)=\delta(x-y)\,\Box^{s}\mathscr{D}_J\;,
\end{equation}
where we remind the reader that $\mathscr{D}_J$ is the traceless-symmetric rank-$J$ derivative. The scalar case $F=\tfrac12\OO^2$ of \cref{sec:O2} is recovered for $\KK(x,y)=\delta(x-y)$ ($s=J=0$). The generalized Hubbard--Stratonovich transform and the saddle-point analysis of \cref{sec:O2} carry over, giving the one-loop functional
\begin{equation}\label{eq:Wlambda_gen}
\begin{split}
  \widetilde W[\JJ,\lambda]={}&W[\widehat\JJ]
  -\tfrac12\int \dd^4x\,\dd^4y\,\lambda^{\mu_1\cdots\mu_J}(x)\,\langle\OO(x)\rangle_{\widehat\JJ}\,
       \KK_{\mu_1\cdots\mu_J}(x,y)\,\langle\OO(y)\rangle_{\widehat\JJ}\\
  &+\tfrac12\,\mathrm{Tr}\log\!\bigl(1+\lambda\KK\mathscr{G}_{\widehat\JJ}\bigr)+\mathcal{O}(N^{-2})\;.
\end{split}
\end{equation} 
The implementation of the functional derivative $\delta/\delta\lambda^{\mu_1\cdots\mu_J}(x)$ at $\JJ,\lambda\to0$ in this generating functional follows \eqref{eq:dJhat}--\eqref{eq:O2half} mutatis mutandis. The tree-level term drops out assuming $\langle\OO\rangle_\beta=0$, and the surviving trace-log piece is the coincident-point trace with the local kernel acting as the momentum-space multiplier $\widetilde\KK(n,k)=\int\dd\tau\,\dd^3x\,e^{i\omega_n\tau+i\bk\cdot\bx}\KK(x,0)$. Accordingly, the bare thermal one-point function takes the form
\begin{equation}\label{eq:gendeform_bare}  \langle\OO\,\Box^{s}\mathscr{D}_J\OO\rangle_\beta^{\rm bare}
=\NN_{\Delta_\OO}\,\frac1\beta\sum_{n\in\mathbb{Z}}\int_{\mathbb{R}^3}\frac{\dd^3k}{(2\pi)^3}\,
  \widetilde\KK(n,k)\,\GG_T(n,k)\;,
\end{equation}
with a momentum integral over $\GG_T$ suitably weighted by the deformation kernel. Equivalently, this is the coincident-point limit of $\Box^{s}\mathscr{D}_J$ acting on the thermal two-point function, the single connected contraction of $\OO\,\Box^{s}\mathscr{D}_J\OO$, which is the full content of the leading-order result.

It is worth elaborating further on the form of the tensorial structure of the integral on the RHS of \eqref{eq:gendeform_bare}. The kernel in~\eqref{eq:gendeform_bare} is the momentum-space image of the local operator
$\Box^{s}\mathscr{D}_J$. With the substitution $\partial\to i\pp$, $\Box^{s}\to(-p_n^2)^{s}$, and
$\mathscr{D}_J\to i^{J}\mathscr{P}_J$, where
$\mathscr{P}_J^{\mu_1\cdots\mu_J}=p_n^{\mu_1}\cdots p_n^{\mu_J}-(\text{traces})$, we obtain for 
even $J$
\begin{equation}\label{eq:Kgen}
  \widetilde\KK(n,k)=(-1)^{s+J/2}\,p_n^{2s}\,\mathscr{P}_J^{\mu_1\cdots\mu_J}\;,\qquad
  p_n^2=\omega_n^2+k^2\;.
\end{equation}
Inserting~\eqref{eq:Kgen} into~\eqref{eq:gendeform_bare} yields the one-point function
\begin{equation}\label{eq:gendeform_tensor}  \langle\OO\,\Box^{s}\mathscr{D}_J^{\mu_1\cdots\mu_J}\OO\rangle_\beta^{\rm bare}
  =(-1)^{s+J/2}\,\NN_{\Delta_\OO}\,\frac1\beta\sum_{n\in\mathbb{Z}}\int_{\mathbb{R}^3}\frac{\dd^3k}{(2\pi)^3}\,
  p_n^{2s}\,\mathscr{P}_J^{\mu_1\cdots\mu_J}\,\GG_T(n,k)\;.
\end{equation}
The integrand of~\eqref{eq:gendeform_tensor} depends on the direction $\hat\bk$ of the spatial momentum only through $\mathscr{P}_J^{\mu_1\cdots\mu_J}(p)$, since the thermal response $\GG_T(n,k)$ is a function of $|\bk|=k$ alone and the thermal state preserves spatial $O(3)$ rotations. In spherical coordinates, $\int\dd^3k=\int_0^\infty\dd k\,k^2\int_{S^2}\dd\Omega_2$,
the angular integral acts non-trivially only on $\mathscr{P}_J$, which can be replaced by its average over the directions $\hat\bk$ at fixed $(\omega_n,k)$. By $O(3)$ invariance this average is built only from the structures that survive integration over $\bk$: the metric $\delta^{\mu\nu}$ and the thermal direction $\hat\tau^\mu\equiv\delta^{\mu0}$ singled out by $\omega_n$. Therefore, being symmetric and traceless, it is proportional to the unique invariant traceless-symmetric rank-$J$ tensor, $\mathcal{Q}_J^{\mu_1\cdots\mu_J}=\delta^{\mu_1 0}\cdots\delta^{\mu_J 0}-(\text{traces})$.
The proportionality constant follows from the zonal harmonic identity (see \cref{app:gegenbauer}): contracting the harmonic (traceless) tensor $\mathscr{P}_J$ with any unit vector $\hat a$ collapses it to a single Gegenbauer polynomial,
\begin{equation}\label{eq:harmonic_id}
  \mathscr{P}_J^{\mu_1\cdots\mu_J}(p)\,\hat a_{\mu_1}\cdots\hat a_{\mu_J}
  =\frac{J!}{2^{J}(\hat\nu)_J}\,p_n^{J}\,C_J^{(\hat\nu)}({\hat p}_n\cdot\hat a)\;.
\end{equation}
Setting $\hat a=\hat\tau$ gives $\hat p_n\cdot\hat\tau=\omega_n/p_n$ and hence $\mathscr{P}_J^{\mu_1\cdots\mu_J}\hat\tau_{\mu_1}\cdots\hat\tau_{\mu_J}=2^{-J}p_n^{J}C_J^{(1)}(\omega_n/p_n)$. The same contraction applied to $\mathcal{Q}_J$ (i.e.~\eqref{eq:harmonic_id} at $p=\hat\tau$) gives $\mathcal{Q}_J^{\mu_1 \cdots\mu_J}\hat\tau_{\mu_1}\cdots\hat\tau_{\mu_J}=2^{-J}C_J^{(1)}(1)=2^{-J}(J+1)$. Their ratio fixes the constant, so under the angular integral
\begin{equation}\label{eq:PJaverage}
  \mathscr{P}_J^{\mu_1\cdots\mu_J}\ \longrightarrow\
  \mathcal{Q}_J^{\mu_1\cdots\mu_J}\,\frac{p_n^{J}}{J+1}\,C_J^{(1)}\!\Bigl(\frac{\omega_n}{p_n}\Bigr)\;.
\end{equation}
Substituting the angular average~\eqref{eq:PJaverage} into~\eqref{eq:gendeform_tensor}, and performing the radial reduction $\int\!\tfrac{\dd^3k}{(2\pi)^3}=\tfrac{1}{2\pi^2}\int_0^\infty\!\dd k\,k^2$,
we obtain
\begin{equation}\label{eq:bgen}
\langle\OO\,\Box^{s}\mathscr{D}_J\OO\rangle_\beta^{\rm bare}
  =\NN_{\Delta_\OO}\,\frac{(-1)^{s+J/2}}{J+1}\,
  \frac{\QQ_J} 
  {2\pi^2\beta}\sum_{n}\int_0^\infty\!\dd k\,k^2\,
  p_n^{2s+J}\,C_J^{(1)}\!\Bigl(\frac{\omega_n}{p_n}\Bigr)\GG_T(n,k)\;.
\end{equation}
This formula generalizes the bare $\OO^2$ expression in \eqref{eq:O2bare} to the general double-twist operator. What remains is to properly regularize and define the divergent momentum integral (in analogy to \eqref{eq:a00_def}) and connect the resulting thermal one-point function to the thermal OPE coefficients $a_{s,J}$. This is the main task of the next section.

\section{The OPE perspective}
\label{sec:equivalence}

The previous section demonstrates a direct link between the thermal one-point functions of the double-twist operators and the thermal response function, and suggests that a proper derivation of this relation involves a point-splitting definition of the double-twist operators $[\OO\OO]_{s,J}\sim\OO\,\Box^{s}\mathscr{D}_J\OO$. In this section, we implement the coincidence limit $|x|\to 0$ of the point-split operators directly inside the (differentiated) thermal two-point function $\Box^{s}\mathscr{D}_J g(\tau, \bx)$ and isolate the contribution of the thermal block of interest. Not only do we recover the bare one-point functions \eqref{eq:bgen}, but we also understand how to properly regularize them and how to isolate the thermal OPE coefficients $a_{s,J}$.

\subsection{Derivatives of thermal blocks}
\label{sec:block_derivatives}

Recall from~\eqref{eq:thermal_OPE} that a single thermal block is the summand $\beta^{-\Delta}\,C_J^{(1)}(\eta)\,|x|^{-2\Delta_\OO+\Delta}$, i.e. a power of $|x|$ times a Gegenbauer polynomial in $\eta=\tau/|x|$. The Laplacian acts simply on such a block,
\begin{equation}\label{eq:box_block}
  \Box\!\left(|x|^{-2\Delta_\OO+\Delta}\,C_J^{(1)}(\eta)\right)
  =\bigl[(\Delta-2\Delta_\OO)(\Delta-2\Delta_\OO+2)-J(J+2)\bigr]\,
  |x|^{-2\Delta_\OO+\Delta-2}\,C_J^{(1)}(\eta)\;,
\end{equation}
lowering the power of $|x|$ by two, while the traceless-symmetric rank-$J$ derivative $\mathscr{D}_J$ lowers it by $J$ and turns the single Gegenbauer into a rank-$J$ tensor built from $\delta^{\mu 0}$, $x^\mu/|x|$ and $\delta^{\mu\nu}$. Acting with the combined order-$(2s+J)$ operator $\Box^{s}\mathscr{D}_J$ on a block of {\em generic} dimension $\Delta$ therefore gives
\begin{equation}\label{eq:block_scaling}
  \Box^{s}\mathscr{D}_J^{\mu_1\cdots\mu_J}\!\left(|x|^{-2\Delta_\OO+\Delta}\,C_J^{(1)}(\eta)\right)
  =|x|^{-2\Delta_\OO+\Delta-2s-J}\times(\text{rank-}J\text{ tensor})(\eta)\;.
\end{equation}
By generic dimension, we refer to dimensions $\Delta$ such that $\Delta-2\Delta_\OO$ is not a non-negative even integer. When the latter holds, the block is a monomial with an integer power in $|x|^2$ and too many derivatives annihilate it instead of producing negative powers.

The exponent that appears on the RHS of Eq.~\eqref{eq:block_scaling} controls the coincidence limit. For generic $\Delta$, blocks with $\Delta>2\Delta_\OO+2s+J$ vanish as $|x|\to0$, those with $\Delta<2\Delta_\OO+2s+J$ diverge, and the marginal value $\Delta=2\Delta_\OO+2s+J$---precisely the double-twist dimension~\eqref{eq:dt_dim} corresponding to $\Box^{s}\mathscr{D}_J^{\mu_1\cdots\mu_J}$---gives a
finite constant. For the latter,
\begin{equation}\label{eq:block_limit}
  \Box^{s}\mathscr{D}_J^{\mu_1\cdots\mu_J}\!\left(|x|^{2s'+J'}\,C_{J'}^{(1)}(\eta)\right)
  \xrightarrow{\;|x|\to0\;}
  \delta_{s,s'}\,\delta_{J,J'}\,\mathcal{A}(s,J)\,\mathcal{Q}_J^{\mu_1\cdots\mu_J}\;,\quad
  \mathcal{A}(s,J)=2^{2s+J}\,s!\,J!\,(J+2)_s\;,
\end{equation}
with $\mathcal{Q}_J$ the invariant tensor of Eq.~\eqref{eq:PJaverage}.\footnote{For $J=0$,
\eqref{eq:box_block} gives $\Box(|x|^{2s})=4s(s+1)|x|^{2s-2}$, so iterating $s$ times yields $\mathcal{A}(s,0)=\prod_{j=1}^{s}4j(j+1)=2^{2s}s!\,(2)_s$. The general $J$ result follows from the analogous action of $\mathscr{D}_J$ and is confirmed by the explicit cases that we discuss below.} Therefore, after the potential subtraction of a finite number of divergent terms, $\Box^{s}\mathscr{D}_J$ acts in the coincidence limit as a projector within the double-twist tower: it annihilates every block whose quantum numbers differ from $(s,J)$, leaving only the contribution of $[\OO\OO]_{s,J}$ itself.

\subsection{Projection and the bare one-point function formula}
\label{sec:block_projection}

Acting on the full expansion~\eqref{eq:OPE4d} with $\Box^{s}\mathscr{D}_J$ and
using~\eqref{eq:block_limit}, the target block $[\OO\OO]_{s,J}$ survives as a constant, while
every other double-twist block is annihilated---the heavier ones by positive powers of
$|x|$ in the coincidence limit, the lighter ones as lower-degree polynomials by the derivatives alone. The only blocks left alongside the
target are the lighter \emph{non}-double-twist contributions ($\Delta<2\Delta_\OO+2s+J$: the
identity and the energy-momentum sector), which are not polynomials and instead diverge,
\begin{align}\label{eq:diff_OPE}
  \Box^{s}\mathscr{D}_J^{\mu_1\cdots\mu_J}\,g(\tau,\bx)
  ={}&\mathcal{A}(s,J)\,\frac{a_{s,J}}{\beta^{2\Delta_\OO+2s+J}}\,\mathcal{Q}_J^{\mu_1\cdots\mu_J}
  \nonumber\\
&+\Box^{s}\mathscr{D}_J^{\mu_1\cdots\mu_J}\!\!\sum_{\Delta<2\Delta_\OO+2s+J}\!\!(\text{lighter ~non-double-twist~blocks})\nonumber\\
  &+(\text{vanishing as }|x|\to0)\;.
\end{align}
The light blocks carry angular dependence, so the bare $|x|\to0$ limit of $\Box^{s}\mathscr{D}_J\,g$ is ambiguous. We make it definite by averaging over the sphere, $\frac{1}{2\pi^2}\int_{S^3}\dd\Omega_3$ (with $\mathrm{Vol}\,S^3=2\pi^2$). This single operation projects in two ways: it sends the scalar $\eta$-dependence onto the spin-zero Gegenbauer polynomial $C_0^{(1)}=1$ (using the orthogonality property~\eqref{eq:gegen_ortho}) and the tensor onto the unique invariant structure $\mathcal{Q}_J$ (the free indices being spectators and the only external vector surviving the average being the thermal direction $\delta^{\mu0}$). The target block, already $\eta$-independent and proportional to $\mathcal{Q}_J$ after differentiation~\eqref{eq:block_limit}, passes through untouched, while each light block contributes only its spin-zero part. Therefore, at finite $|x|$ the average leads to the identity
\begin{equation}\label{eq:projected_OPE}
  \frac{1}{2\pi^2}\int_{S^3}\!\dd\Omega_3\,\Box^{s}\mathscr{D}_J^{\mu_1\cdots\mu_J}\,g
  =\mathcal{A}(s,J)\,\frac{a_{s,J}}{\beta^{2\Delta_\OO+2s+J}}\,\mathcal{Q}_J^{\mu_1\cdots\mu_J}
  +\mathcal{X}_T^{\mu_1\cdots\mu_J}(|x|)
  +(\text{vanishing as }|x|\to0)\;,
\end{equation}
where the target term is $|x|$-independent, the heavier double-twists make up the vanishing tail, and the lighter (identity and multi-stress) blocks collect into the remainder
\begin{equation}\label{eq:XT}
  \mathcal{X}_T^{\mu_1\cdots\mu_J}(|x|)\coloneqq
  \frac{1}{2\pi^2}\int_{S^3}\!\dd\Omega_3\,\Box^{s}\mathscr{D}_J^{\mu_1\cdots\mu_J}\!\!
  \sum_{4m<2\Delta_\OO+2s+J}\;\sum_{\substack{\ell=0\\ \rm even}}^{2m}
  \frac{a^{(m)}_\ell}{\beta^{4m}}\,C_\ell^{(1)}(\eta)\,|x|^{-2\Delta_\OO+4m}\;,
\end{equation}
which grows without bound as $|x|\to0$.

Now, we notice that the LHS of Eq.~\eqref{eq:projected_OPE} can be computed independently from the momentum-space representation~\eqref{eq:gGT} of the two-point function, with $p_n=(\omega_n,\boldsymbol{k})$. The operator $\Box^{s}\mathscr{D}_J$ multiplies the plane wave by its momentum-space symbol $\widetilde\KK(\pp)$ in~\eqref{eq:Kgen},
\begin{equation}\label{eq:diffg_momentum}
\Box^{s}\mathscr{D}_J^{\mu_1\cdots\mu_J}\,g(\tau,\boldsymbol{x})
  =\frac{\NN_{\Delta_\OO}}{\beta}\sum_n\int\frac{\dd^3k}{(2\pi)^3}\,  (-1)^{s+J/2}\,p_n^{2s}\,\mathscr{P}_J^{\mu_1\cdots\mu_J}\,e^{ip_n\cdot x}\,\GG_T(n,k)\;.
\end{equation}
The spatial average~\eqref{eq:PJaverage} replaces
$\mathscr{P}_J^{\mu_1\cdots\mu_J}\to\mathcal{Q}_J^{\mu_1\cdots\mu_J}\,p_n^{J}C_J^{(1)}(\omega_n/p_n)/(J+1)$, leaving an integrand independent of the direction of the spatial momentum vector $\boldsymbol{k}$, while the remaining average over the direction $\hat x=x/|x|\in S^3$ acts only on the plane wave. The hyperspherical machinery of Appendix~\ref{app:gegenbauer}---the Rayleigh expansion~\eqref{eq:rayleigh} of $e^{ip_n\cdot x}$ and the harmonic orthogonality~\eqref{eq:gegen_ortho}---collapses $\tfrac{1}{2\pi^2}\int_{S^3}\dd\Omega_3\,e^{ip_n\cdot x}$ to a single radial integral with a Bessel kernel, the spin-zero case of the projection identity~\eqref{eq:gegen_fourier_proj_final}. Together these give
\begin{align}\label{eq:projected_momentum}
  \frac{1}{2\pi^2}&\int_{S^3}\!\dd\Omega_3\,\Box^{s}\mathscr{D}_J^{\mu_1\cdots\mu_J}\,g(\tau,\boldsymbol{x})=\cr
  &\NN_{\Delta_\OO}\,\frac{(-1)^{s+J/2}}{J+1}\,\mathcal{Q}_J^{\mu_1\cdots\mu_J}\,
  \frac{1}{2\pi^2\beta}\sum_n\int_0^\infty\!\dd k\,k^2\,p_n^{2s+J}\,
  C_J^{(1)}\!\Big(\frac{\omega_n}{p_n}\Big)\,\frac{2 J_1(p_n|x|)}{p_n|x|}\,\GG_T(n,k)\;.
\end{align}

Finally, let us examine the behavior of this expression in the coincidence limit. On the LHS we obtain the bare double-twist one-point function $\langle\OO\,\Box^{s}\mathscr{D}_J^{\mu_1\cdots\mu_J}\OO\rangle_\beta^{\rm bare}$. On the RHS, the Bessel kernel $J_1(\pp|x|)/(\pp|x|/2)$ keeps the integral finite at nonzero $|x|$ and tends to unity as $|x|\to0$ recovering the main result \eqref{eq:bgen} in \cref{sec:deformations}. Moreover, by equating the two evaluations~\eqref{eq:projected_OPE}
and~\eqref{eq:projected_momentum} we learn that 
the $|x|\to0$ divergence is captured by the remainder $\mathcal{X}_T$. This allows us to obtain the regularized expressions
\begin{align}\label{eq:equiv_master}
\langle\OO\,\Box^{s}\mathscr{D}_J^{\mu_1\cdots\mu_J}\OO\rangle_\beta^{\rm ren}
&= \langle\OO\,\Box^{s}\mathscr{D}_J^{\mu_1\cdots\mu_J}\OO\rangle_\beta^{\rm bare} - \mathcal{X}_T^{\mu_1\cdots\mu_J}
\cr
&=\NN_{\Delta_\OO}\,\frac{(-1)^{s+J/2}}{J+1}\,
  \frac{\QQ_J^{\mu_1\cdots\mu_J}} {2\pi^2\beta}\sum_{n}\int_0^\infty\!\dd k\,k^2\,
  p_n^{2s+J}\,C_J^{(1)}\!\Bigl(\frac{\omega_n}{p_n}\Bigr)\GG_T(n,k) - 
  \mathcal{X}_T^{\mu_1\cdots\mu_J}  
\cr
&=\mathcal{A}(s,J)\,\frac{a_{s,J}}{\beta^{2\Delta_\OO+2s+J}}\,\mathcal{Q}_J^{\mu_1\cdots\mu_J}\;,
\end{align}
which can be used to obtain the thermal OPE coefficients $a_{s,J}$ from the thermal response function $\GG_T(n,k)$ after suitably subtracting the divergent identity/multi-stress-tensor contributions in $\mathcal{X}_T$.

\subsection{Specific examples}

It is instructive to explain how everything works in the three lightest double-twist operators $[\OO\OO]_{0,0}$, $[\OO\OO]_{1,0}$, $[\OO\OO]_{0,2}$ and record the resulting expressions for the corresponding thermal OPE coefficients in preparation for the numerical evaluation of \cref{sec:regularization}.

\paragraph{$\boldsymbol{s=0,\,J=0}$.} 
In this case, $\mathcal{Q}_0=1$, $\mathcal{A}(0,0)=1$, and the only lighter block that contributes to the divergent part is the identity. Hence, $\mathcal{X}_T=\NN_{\Delta_\OO}\int\tfrac{\dd^4p}{(2\pi)^4}\GG_0(p)$ is the coincident-point $T=0$ two-point function and~\eqref{eq:equiv_master} reduces to the renormalized scalar double-trace thermal one-point function~\eqref{eq:a00_def}.

\paragraph{$\boldsymbol{s=1,\,J=0}$.} 
This is again a scalar case and $\mathcal{A}(1,0)=8$, with the identity and the energy-momentum tensor below threshold. The spin-two energy-momentum tensor term keeps its $C_2^{(1)}(\eta)$ dependence under $\Box$ and is annihilated by the spin-zero projection, so only the identity contributes. As a result, 
\begin{equation}
\mathcal{X}_T=4\Delta_\OO(\Delta_\OO-1)|x|^{-2\Delta_\OO-2}\big|_{|x|\to0}
=-\NN_{\Delta_\OO}\int\tfrac{\dd^4p}{(2\pi)^4}p^2\GG_0(p)~.
\end{equation}

\paragraph{$\boldsymbol{s=0,\,J=2}$.} 
This is a case of a spin-two double-twist operator. We have $\mathcal{Q}_2^{\mu\nu}=\delta^{\mu0}\delta^{\nu0}-\tfrac14\delta^{\mu\nu}$, and $\mathcal{A}(0,2)=8$. Now $\mathscr{D}_2^{\mu\nu}$ acting on the identity gives a traceless $x^\mu x^\nu/|x|^2$ structure that averages to zero on $S^3$, so the surviving divergence comes from the contribution of the energy-momentum tensor block
\begin{align}
\mathcal{X}_T^{\mu\nu}
&=
\frac{a^{(1)}_2}{\beta^4}
\frac{4}{3}
(\Delta_{\mathcal O}-3)(\Delta_{\mathcal O}-4)
|x|^{-2\Delta_{\mathcal O}+2}\big|_{|x|\to0}\,
\mathcal{Q}_2^{\mu\nu}
\nonumber\\
&=
-\frac{a^{(1)}_2}{\beta^4}
\frac{8\pi^2}{3\nu}
(\Delta_{\mathcal O}-3)(\Delta_{\mathcal O}-4)
\int_{\mathbb{R}^4}
\frac{\dd^4  p}{(2\pi)^4}\,
p^{-2}\mathcal{G}_0(p)\,
\mathcal{Q}_2^{\mu\nu}
~.
\end{align}
Unlike the previous scalar cases, $a_{0,2}$
requires a genuine energy-momentum subtraction. Only same-spin blocks survive the projection.

Applying~\eqref{eq:equiv_master} to these examples gives the explicit double-twist coefficient relations:
\begin{align}\label{eq:a_explicit1}  a_{0,0}&=\NN_{\Delta_\OO}\,\beta^{2\Delta_\OO}\left[\frac{1}{2\pi^2\beta}
    \sum_{n\in\mathbb{Z}}\int_0^\infty\!\dd k\,k^2\,\GG_T(n,k)
    -\int_{\mathbb{R}^4}\frac{\dd^4p}{(2\pi)^4}\,\GG_0(p)\right],\\
    \label{eq:a_explicit2}
  a_{1,0}&=-\frac{\NN_{\Delta_\OO}\,\beta^{2\Delta_\OO+2}}{8}\left[\frac{1}{2\pi^2\beta}
    \sum_{n\in\mathbb{Z}}\int_0^\infty\!\dd k\,k^2\,(\omega_n^2+k^2)\,\GG_T(n,k)
    -\int_{\mathbb{R}^4}\frac{\dd^4p}{(2\pi)^4}\,p^2\,\GG_0(p)\right],\\
    \label{eq:a_explicit3}
  a_{0,2}&=-\frac{\NN_{\Delta_\OO}\,\beta^{2\Delta_\OO+2}}{8}\Bigg[\frac{1}{2\pi^2\beta}
    \sum_{n\in\mathbb{Z}}\int_0^\infty\!\dd k\,k^2\,\frac{3\omega_n^2-k^2}{3}\,\GG_T(n,k)\nonumber\\
    &\qquad \qquad\qquad \qquad \qquad
    -
\frac{a^{(1)}_2}{\beta^4}
\frac{16}{3}
(\Delta_{\mathcal O}-1)(\Delta_{\mathcal O}-2)
(\Delta_{\mathcal O}-3)(\Delta_{\mathcal O}-4)
\int_{\mathbb R^4}
\frac{\dd^4 p}{(2\pi)^4}\,
p^{-2}\mathcal G_0(p)\,
    \Bigg]~.
\end{align}

For $a_{0,0}$ and $a_{1,0}$ the (identity) subtraction involves the $T=0$ 
correlator, weighted by $1$ and $p^2$ respectively. For $a_{0,2}$ the identity value vanishes by
the restoration of $O(4)$ invariance at $T=0$, and the subtraction involves instead the
energy-momentum block, with $a^{(1)}_2=\Delta_\OO\,\pi^4/120$ the universal energy-momentum thermal one-point coefficient~\cite{Fitzpatrick:2019zqz}. 

\section{Regularization and evaluation}
\label{sec:regularization}

The explicit coefficients~\eqref{eq:a_explicit1}-\eqref{eq:a_explicit3} (and their higher $(s,J)$ generalizations) are written as differences between individually divergent quantities. In this section, we implement a concrete evaluation procedure for such expressions and use it to compute numerically the values of the thermal OPE coefficients
$a_{0,0}$, $a_{1,0}$ and $a_{0,2}$, comparing where possible with available results in the literature.

\subsection{A window-subtraction scheme}
\label{sec:window}

The computation of the thermal coefficients $a_{s,J}$ in \eqref{eq:equiv_master}
\begin{align}
\label{winaa}
a_{s,J}\,\mathcal{Q}_J^{\mu_1\cdots\mu_J}&=
\frac{\beta^{2\Delta_\OO+2s+J}}{\mathcal{A}(s,J)} \times
\\
&\times \bigg[\NN_{\Delta_\OO}\,\frac{(-1)^{s+J/2}}{(J+1)}\,
  \frac{\QQ_J^{\mu_1\cdots\mu_J}} {2\pi^2 \beta}\sum_{n}\int_0^\infty\!\dd k\,k^2\,
  \pp^{2s+J}\,C_J^{(1)}\!\Bigl(\frac{\omega_n}{\pp}\Bigr)\GG_T(n,k) - 
  \mathcal{X}_T^{\mu_1\cdots\mu_J} \bigg] 
  \nonumber
\end{align}
involves the careful cancellation of ultraviolet divergences between the two terms on the RHS. To make the computation of these expressions more amenable to explicit numerical computation, we introduce the following scheme. 

First, we need to identify the precise source of the ultraviolet divergence in the radial integrals 
\begin{equation}\label{eq:Wdef}
  \mathcal{W}^{(s,J)}
  =\frac{1}{2\pi^2\beta}\sum_{n\in\mathbb{Z}}\int_0^\infty\!\dd k\,k^2\,  \pp^{2s+J}\,C_J^{(1)}\!\Bigl(\frac{\omega_n}{\pp}\Bigr)\,\GG_T(n,k)\;,
\end{equation}
where, as always, $\pp^2\coloneqq\omega_n^2+k^2$.
At large $k$, where $\GG_T\to\GG_0\sim k^{2\Delta_\OO-4}$, the integrand grows as $k^{2\Delta_\OO+2s+J-2}$ capturing the leading source of the divergence. In general, there are additional subleading sources of divergence. Since the large-momentum expansion of $\GG_T$ is governed by the OPE, the corresponding asymptotics are fixed by the operators with power-law momentum tails. Crucially, every double-twist block has an exponentially small momentum tail, and the power-law growth comes entirely from the identity and the multi-stress tower (see Appendix~\ref{app:highmom}). For example, the large-momentum asymptotics of $\GG_T$ corresponding to the identity and the energy-momentum contributions are
\begin{equation}\label{eq:GTasympt}  \GG_T(n,k)=\GG_0(p_n)\left[1+16\,a^{(1)}_2\,\frac{\Gamma(\Delta_\OO)}{\Gamma(\Delta_\OO-4)}\,
  \frac{3\omega_n^2-k^2}{\beta^4 \pp^6}+\mathcal{O}(\pp^{-8})\right]\;.
\end{equation}
The constant $a^{(1)}_2$ is the energy-momentum tensor thermal one-point coefficient. Since each multi-stress-tensor term lowers the large-$k$ growth by an additional inverse power $k^{-4}$, the number of terms in the large-$p$ expansion that contribute to the full divergence of $\WW^{(s,J)}$ (including the Matsubara sum, which leads effectively to a four-dimensional momentum integral) is captured by the first $M_{s,J}\coloneqq\lceil(2\Delta_\OO+2s+J)/4\rceil$ terms in the large-$p$ expansion of $\GG_T$. To be concrete, let us express the large-$p$ divergence-generating part of $\GG_T$ as a finite sum of the form
\begin{equation}
\label{winsumaa}
\GG^{(\rm div)}_T(n,k) = \sum_{m=0}^{M_{s,J}-1} \GG_T^{(m)}(n,k)
~.
\end{equation}

The individual contributions $\GG_T^{(m)}(n,k)$ capture well the ultraviolet behavior of $\GG_T$ but are badly inappropriate in the infrared. Since they involve inverse powers of the momentum, inserting them in the momentum integrals $\sum_n \int_0^\infty \dd k$ can lead to infrared divergences. This problem can be mitigated by localizing each $\GG_T^{(m)}(n,k)$ within a window around the UV with a dressing function $w_m(\mu,p)$ that approaches unity in the ultraviolet and an appropriate positive power in the infrared,
\begin{equation}\label{eq:window_props}
w_m(\mu,p)\xrightarrow{\;p\to\infty\;}1\;,\qquad
  w_m(\mu,p)\xrightarrow{\;p\to0\;}(p/\mu)^q+\mathcal{O}((p/\mu)^{>q})\;.
\end{equation}
The scale $\mu$ is arbitrary and drops out at the end of the computation. An example of the type of dressing we will choose later is
\begin{equation}\label{eq:window}
  w(\mu,p)=1-e^{-p^2/\mu^2}\;.
\end{equation}
The corresponding UV-localized version of Eq.~\eqref{winsumaa} is therefore
\begin{equation}
\label{winsumab}
\GG^{(\rm div,win)}_T(n,k) = \sum_{m=0}^{M_{s,J}-1} w_m(\mu,\pp)\GG_T^{(m)}(n,k)  
~.
\end{equation}
Subtracting and adding back $\GG^{(\rm div,win)}_T(n,k)$ inside $\WW^{(s,J)}$ yields an equivalent expression for the double-twist coefficients in \eqref{winaa} of the form
\begin{align}
\label{winab}
&a_{s,J}=
\frac{\beta^{2\Delta_\OO+2s+J}}{\mathcal{A}(s,J)} \times
\nonumber\\
&\times \bigg\{\NN_{\Delta_\OO}\,\frac{(-1)^{s+J/2}}{(J+1)}\,
  \frac{1}
  {2\pi^2 \beta}\sum_{n}\int_0^\infty\!\dd k\,k^2\,
  \pp^{2s+J}\,C_J^{(1)}\!\Bigl(\frac{\omega_n}{\pp}\Bigr)\bigg(\GG_T(n,k) -\sum_{m=0}^{M_{s,J}-1}w_m \GG_T^{(m)}(n,k) \bigg) 
\nonumber\\
&+
\NN_{\Delta_\OO}\,\frac{(-1)^{s+J/2}}{(J+1)}\,
  \frac{1}
  {2\pi^2 \beta}\sum_{n}\int_0^\infty\!\dd k\,k^2\,
  \pp^{2s+J}\,C_J^{(1)}\!\Bigl(\frac{\omega_n}{\pp}\Bigr) \sum_{m=0}^{M_{s,J}-1}w_m \GG_T^{(m)}(n,k)
  - \mathcal{X}_T \bigg\}~,
\end{align}
where we defined $\XX_T$ by the $|x|\to 0$ limit $\XX_T^{\mu_1\cdots \mu_J} \to \XX_T \mathcal{Q}_J^{\mu_1\cdots\mu_J}$.

Now the second line is a finite residual that can be computed efficiently numerically. The regulated cancellation of UV infinities is relegated to the third line with one important advantage: the first term involving the add-back functions $\GG_T^{(m)}$ can now be computed analytically, facilitating a more practical evaluation of the finite leftover. 

We note that in this subtract-and-add-back scheme, $M_{s,J}$ is the minimum number of terms that should be incorporated from the large-$\pp$ asymptotics of $\GG_T$. Higher-order terms can also be incorporated freely. They change the finite residual on the second line of \eqref{winab} by a finite amount and typically improve the numerical accuracy of the computation. Since the coefficients of \eqref{eq:GT_asym_general} grow factorially, the scale $\mu$ of the
dressing function of the $m$-th term should be taken to grow with $m$.

We proceed to demonstrate explicitly how this scheme is implemented on the first few double-twist coefficients in the following subsections.

\subsection{A detailed example: $a_{0,0}$}
To demonstrate the practical aspects of the calculation scheme introduced in the previous subsection, we consider the first double-twist operator $[\OO \OO]_{0,0}$. The corresponding thermal coefficient is given by \eqref{eq:a_explicit1}. To regulate UV divergences, we subtract and add back the asymptotic \eqref{eq:GTasympt}\footnote{The minimum number of needed asymptotic subtractions here is $\lceil\Delta_\OO/2 \rceil$. Since we are interested in the region $\Delta_\OO\in(1,2)$ this can be satisfied by subtracting only the identity, however in such a case the sum/integral would converge only like $\sim p^{2\Delta_\OO-4}$.
To ensure a stronger convergence, we additionally subtract the next asymptotic term, i.e. the energy-momentum tensor.} arriving at the expression
\begin{align}\label{eq:a00_improved_subtraction}
a_{0,0}
=
\NN_{\Delta_\OO}\beta^{2\Delta_\OO}
\Bigg[
&
\frac{1}{\beta}
\sum_{n\in\mathbb Z}
\int_{\mathbb R^3}
\frac{\dd^3k}{(2\pi)^3}
\left(
\GG_T(n,k)-\GG_0(\pp)
-
\frac{16a_2^{(1)}\Gamma(\Delta_\OO)}{\Gamma(\Delta_\OO-4)}
w(\mu,\pp)\,
\GG_0(\pp)
\frac{3\omega_n^2-k^2}{\beta^4\pp^6}
\right)
\nonumber\\
&+
\left(
\frac{1}{\beta}
\sum_{n\in\mathbb Z}
\int_{\mathbb R^3}
\frac{\dd^3k}{(2\pi)^3}
\GG_0(\pp)
-
\int_{\mathbb R^4}
\frac{\dd^4p}{(2\pi)^4}
\GG_0(p)
\right)
\\
&+
16a_2^{(1)}
\frac{\Gamma(\Delta_\OO)}{\Gamma(\Delta_\OO-4)}
\frac{1}{\beta^5}
\sum_{n\in\mathbb Z}
\int_{\mathbb R^3}
\frac{\dd^3k}{(2\pi)^3}
w(\mu,\pp)\,
\GG_0(\pp)
\frac{3\omega_n^2-k^2}{\pp^6}
\Bigg]~.\nonumber
\end{align}
For this example, we have chosen the dressing function
\begin{equation}\label{eq:a00_window}
    w(\mu,p)=\left(1-e^{-p^2/\mu^2}\right)^2~,
\end{equation}
which vanishes like $p^4$ for small momenta and, as we shall see below, allows for a relatively easy handling of the add-backs through Gaussian integration.

As we noted previously, the term on the first line of \eqref{eq:a00_improved_subtraction} is convergent and can easily be handled numerically. The term on the second line can be related to the corresponding GFF datum $a_{0,0}^{(\rm GFF)} = 2\zeta(2\Delta_{\OO})$\cite{Iliesiu:2018fao}, since from equations \eqref{eq:thermal_OPE} and \eqref{eq:gGT} we can see that
\begin{equation}\label{eq:a00limit}
    \beta^{2\Delta_\OO}\lim_{|x|\to 0^+}{\left(g_{\rm GFF}(\tau,x) - |x|^{-2\Delta_\OO}\right)}\equiv a_{0,0}^{(\rm GFF)}\; ,
\end{equation}
when comparing to the thermal OPE \eqref{eq:thermal_OPE} of the GFF theory. The term in the last line is the energy-momentum tensor add-back
\begin{align}
\label{b00_a}
\mathcal B_{0,0}(\mu)
\coloneqq
\NN_{\Delta_\OO}\,
16a_2^{(1)}
\frac{\Gamma(\Delta_\OO)}{\Gamma(\Delta_\OO-4)}
\frac{1}{\beta^5}
\sum_{n\in\mathbb Z}
\int_{\mathbb R^3}
\frac{\dd^3k}{(2\pi)^3}
\,
\GG_0(\pp)\, \left(1-e^{-\pp^2/\mu^2}\right)^2\,
\frac{3\omega_n^2-k^2}{\pp^6}.
\end{align}
Substituting the zero-temperature response \eqref{eq:G0}, we obtain
\begin{align}
\BB_{0,0}(\mu)=
\SS_{\Delta_{\OO}}
\frac{1}{\beta}
\sum_{n\in\mathbb Z}
\int_0^\infty \dd k\;
k^2
\left(1-e^{-\pp^2/\mu^2}\right)^2
\pp^{2\Delta_\OO-10}
(3\omega_n^2-k^2),
\end{align}
where we defined the coefficient,
\begin{equation}\label{eq:S_Delta}
    \SS_{\Delta_\OO} \coloneqq 2 \nu \,\NN_{\Delta_\OO} \frac{16 a^{(1)}_2 \Gamma(\Delta_\OO)\Gamma(2-\Delta_\OO)}{\pi^22^{2\Delta_\OO-3}\beta^4 \Gamma(\Delta_\OO-4)\Gamma(\Delta_\OO-2)} = 128\frac{a^{(1)}_2}{4^{\Delta_\OO }\beta^4} \frac{\Gamma(2-\Delta_\OO)}{\Gamma(\Delta_\OO-4)} \; .
\end{equation}
To evaluate the sum/integral in \eqref{b00_a}, we use the Schwinger parametrization,
\begin{equation}\label{eq:schwinger}
p^{2\Delta_\OO-10}
=
\frac{1}{\Gamma(5-\Delta_\OO)}
\int_0^\infty \dd s\;
s^{4-\Delta_\OO}e^{-sp^2},
\end{equation}
and after swapping the order of integration with the thermal sum and integral, we arrive at
\begin{align}\label{eq:a00_addback_wtr_I}
\mathcal B_{0,0}(\mu)
=
\frac{\SS_{\Delta_\OO}}{\Gamma(5-\Delta_\OO)}
\int_0^\infty \dd s\;
s^{4-\Delta_\OO}
\frac{1}{\beta}
\sum_{n\in\mathbb Z}
\left[
I_n^{(1)}(s)-2I_n^{(1)}(s+\mu^{-2})+I_n^{(1)}(s+2\mu^{-2})
\right],
\end{align}
where we defined the function
\begin{equation}\label{eq:I00}
    I_n^{(r)}(\alpha) \coloneqq \int_0^\infty \dd k \;e^{-\alpha \pp^2} k^2 (3\omega^2_n-k^2)^r
    = \frac{\sqrt{\pi}}{4}\, e^{-\alpha\omega_n^2} \sum_{j=0}^{r} \binom{r}{j} (3\omega_n^2)^{r-j} \frac{(-1)^j (2j+1)!!}{2^j\, \alpha^{\,j+3/2}} \; .
\end{equation}
At this point we use the Poisson resummation identity 
\begin{equation}\label{eq:poisson}
  \frac1\beta\sum_{n\in\mathbb{Z}}e^{-\alpha\omega_n^2}
  =\frac{1}{\sqrt{4\pi\alpha}}\sum_{m\in\mathbb{Z}}e^{-m^2\beta^2/4\alpha}\;,
\end{equation}
to reformulate the sums over $I_n$ as
\begin{equation}\label{eq:I10_resummed} 
    \frac{1}{\beta}\sum_{n\in \mathbb Z}I_n(\alpha) = -\frac{3\beta^2}{16 \alpha^4} \sum_{m=1}^\infty m^2e^{-\frac{m^2\beta^2}{4\alpha}} \;.
\end{equation}
which exhibit faster convergence properties. Inserting this representation into the expression \eqref{eq:a00_addback_wtr_I}, and using the integral representation,
\begin{equation}
    {}_1F_1(a;b;z)=\frac{\Gamma(b)}{\Gamma(a)\Gamma(b-a)} \int_{0}^1 \dd t\;t^{a-1}(1-t)^{b-a-1}\,e^{zt} \;~,
\end{equation}
we obtain
\begin{align}
\mathcal B_{0,0}(\mu)
=
\frac{24a_2^{(1)}}{4^{\Delta_\OO}\beta^2}
\sum_{m=1}^{\infty}m^2
\Bigg[
&
\left(\frac{m^2\beta^2}{4}\right)^{1-\Delta_\OO}\nonumber
\\
&-
2\mu^{2\Delta_\OO-2}
\frac{\Gamma(5-\Delta_\OO)}{6}
{}_1F_1\left(
\Delta_\OO-1;4;-\frac{m^2\beta^2\mu^2}{4}
\right)\nonumber
\\
&+
2^{1-\Delta_\OO}\mu^{2\Delta_\OO-2}
\frac{\Gamma(5-\Delta_\OO)}{6}
{}_1F_1\left(
\Delta_\OO-1;4;-\frac{m^2\beta^2\mu^2}{8}
\right)
\Bigg],
\end{align}
Consequently, putting everything together we can write
\begin{align}\label{eq:final_a00}
    a_{0,0}
    =    
        \RR_0^{(0,0)}(\mu)
        +
        2\sum_{n=1}^{\infty}\mathcal R_n^{(0,0)}(\mu)
    +
    a_{0,0}^{\rm (GFF)}
    +
    \beta^{2\Delta_\OO} \mathcal B_{0,0}(\mu)\;,
\end{align}
with finite terms
\begin{align}
    \mathcal R_n^{(0,0)}(\mu)
    \coloneqq\frac{\mathcal N_{\Delta_\OO}}{2\pi^2\beta}
    \beta^{2\Delta_{\OO}}&
    \int_0^\infty \dd k\; k^2
    \bigg[
        \GG_T(n,k)
        -
        \GG_0(\pp)
    \nonumber\\
    &- 16\,a_2^{(1)}\frac{\Gamma(\Delta_\OO)}{\Gamma(\Delta_\OO-4)}
    \left(1-e^{-\pp^2/\mu^2}\right)^2
    \GG_0(p_n)
    \frac{3\omega_n^2-k^2}{\beta^4\pp^6}
    \bigg] \;
\end{align}
that form convergent series that can be evaluated by truncation numerically. Explicit numerical results will be presented in \cref{sec:results}.

\subsection{Two more examples: $a_{1,0}$ and $ a_{0,2}$}

In this subsection, we apply a similar procedure to calculate the thermal coefficients of the next order of double-trace data, $[\OO\OO]_{1,0}$ and $[\OO\OO]_{0,2}$. The key features of the derivation remain the same, but certain details change. 

\subsubsection{$a_{1,0}$}
\label{a10}

For the scalar $[\OO\OO]_{1,0}$, the thermal coefficient is given by \cref{eq:a_explicit2}. Subtracting the same asymptotic terms as in the case of $a_{0,0}$, we write
\begin{align}\label{eq:a10_window_general}
    a_{1,0} &= -\NN_{\Delta_{\OO}} \frac{\beta^{2\Delta_\OO+2}}{8}\Bigg[\nonumber \\ 
    &\frac{1}{\beta}\sum_{n\in \mathbb{Z}}\int_{\mathbb{R}^3} \frac{\dd^3 k}{(2\pi)^3}  (\omega_n^2+k^2)\left(\GG_T(n,k) -\GG_0(\pp) -  16\,a_{2}^{(1)}\frac{\Gamma(\Delta_\OO)}{\Gamma(\Delta_\OO-4)}w(\mu,\pp)\;\GG_0(\pp)\frac{3\omega_n^2-k^2}{\beta^4\pp^6} \right) \nonumber  \\ \nonumber &
    + \frac{1}{\beta}\sum_{n\in \mathbb{Z}}\int_{\mathbb{R}^3} \frac{\dd^3 k}{(2\pi)^3}  \pp^2 \GG_0(\pp)  - \int_{\mathbb{R}^4} \frac{\dd^4 p}{(2\pi)^4} p^2 \GG_0(p) \\
    &+ 16\,\frac{a_{2}^{(1)}}{\beta^4}\frac{\Gamma(\Delta_\OO)}{\Gamma(\Delta_\OO-4)}\frac{1}{\beta}\sum_{n\in \mathbb{Z}}\int_{\mathbb{R}^3} \frac{\dd^3 k}{(2\pi)^3}  w(\mu,\pp)\;\GG_0(\pp)\frac{3\omega_n^2-k^2}{\pp^4} 
     \Bigg ] \; .
\end{align}
In this case, the sum/integral involves an extra power of $p^2_n$ compared to \eqref{eq:a00_improved_subtraction}
and therefore, it is enough to choose the dressing function simply as
\begin{equation}
\label{dress1}
    w(\mu,p)=1-e^{-p^2/\mu^2}~.
\end{equation}
Like before, the 
expression in the second line is automatically convergent and suitable for direct numerical evaluation. The combination of terms in the third line can, again, be related to the corresponding GFF datum $a_{1,0}^{\rm (GFF)}=\Delta_\OO \,(\Delta_\OO-1)\zeta(2\Delta_\OO+2)$.

The energy-momentum add-back term in the fourth line is now defined as
\begin{equation}
    \mathcal B_{1,0}(\mu)\coloneqq-2\NN_{\Delta_\OO} \frac{a_2^{(1)}}{\beta^4}\frac{\Gamma(\Delta_\OO)}{\Gamma(\Delta_\OO-4)}\frac{1}{\beta}\sum_{n\in \mathbb{Z}}\int_{\mathbb{R}^3} \frac{\dd^3 k}{(2\pi)^3} (1-e^{-\frac{\pp^2}{\mu^2}})\;\GG_0(\pp)\frac{3\omega_n^2-k^2}{\pp^4} \; 
\end{equation}
\noindent
and repeating the steps of the previous subsection, we can recast it as 
\begin{align}
    \mathcal B_{1,0}(\mu)=\frac{3a_2^{(1)}}{4^{\Delta_\OO} \beta^2}(\Delta_\OO-4)(\Delta_\OO-1)&\sum_{m=1}^\infty m^2\bigg[\left(\frac{2}{m\beta}\right)^{2\Delta_\OO} 
    \nonumber\\
    &-\frac{\mu^{2\Delta_\OO}\Gamma(4-\Delta_\OO)}{6} {}_1 F_1 \left(\Delta_\OO;4;-\frac{m^2 \beta^2 \mu^2}{4}\right)\bigg]~.
\end{align}
Putting everything together, we obtain
\begin{equation}\label{eq:final_a10}
    a_{1,0} = \mathcal R^{(1,0)}_0 (\mu) + 2\sum_{n=1}^{\infty} \mathcal R^{(1,0)}_n (\mu)  + a_{1,0}^{\rm (GFF)} + \beta^{2\Delta_\OO+2} \mathcal B_{1,0}(\mu)
\end{equation}
with
\begin{align}
    \mathcal R_n^{(1,0)}(\mu) = -\frac{\mathcal{N}_{\Delta_\OO}}{2\pi^2\beta} \frac{\beta^{2\Delta_\OO+2}}{8}\int_0^{\infty} \dd k\;k^2 \pp^2\Bigg [\GG_T(n,k)-\GG_0(\pp) \cr - 16\,a_2^{(1)}\frac{\Gamma(\Delta_\OO)}{\Gamma(\Delta_\OO-4)}(1-e^{-\pp^2/\mu^2})\;&\GG_0(\pp)\frac{3\omega_n^2-k^2}{\beta^4\pp^6} \Bigg]~.
\end{align}

\subsubsection{$a_{0,2}$}
\label{a02}

For the spin-2 operator $[\OO\OO]_{0,2}$ we can recast the expression in Eq.~\eqref{eq:a_explicit3} in the following form
\begin{align}\label{eq:a02_window_general}
    a_{0,2} &= -\NN_{\Delta_{\OO}} \frac{\beta^{2\Delta_\OO+2}}{8}\Bigg[\nonumber \\ 
    &\frac{1}{\beta}\sum_{n\in \mathbb{Z}}\int_{\mathbb{R}^3} \frac{\dd^3 k}{(2\pi)^3}  \frac{3\omega_n^2-k^2}{3}\left(\GG_T(n,k) -\GG_0(\pp) -  16\,a_{2}^{(1)}\frac{\Gamma(\Delta_\OO)}{\Gamma(\Delta_\OO-4)}w(\mu,\pp)\;\GG_0(\pp)\frac{3\omega_n^2-k^2}{\beta^4\pp^6} \right) \nonumber  \\ \nonumber &
    + \frac{1}{\beta}\sum_{n\in \mathbb{Z}}\int_{\mathbb{R}^3} \frac{\dd^3 k}{(2\pi)^3}  \frac{3\omega_n^2-k^2}{3}\GG_0(\pp) 
    \\
    &+ \frac{16}{3}\,\frac{a_{2}^{(1)}}{\beta^4}\frac{\Gamma(\Delta_\OO)}{\Gamma(\Delta_\OO-4)}\frac{1}{\beta}\sum_{n\in \mathbb{Z}}\int_{\mathbb{R}^3} \frac{\dd^3 k}{(2\pi)^3}  w(\mu,\pp)\;\GG_0(\pp)\frac{(3\omega_n^2-k^2)^2}{\pp^6}
    \nonumber\\
    & -
\frac{16}{3}\frac{a^{(1)}_2}{\beta^4}
\frac{\Gamma(\Delta_\OO)}{\Gamma(\Delta_\OO-4)}
\int_{\mathbb R^4}
\frac{\dd^4 p}{(2\pi)^4}\,
p^{-2}\mathcal G_0(p)\,
     \Bigg ] \; .
\end{align}
Similar to $a_{1,0}$ we can again choose the dressing function \eqref{dress1}, resulting into a second line that is automatically finite and can be computed numerically. Unlike $a_{1,0}$, however, it is straightforward to check that the third line is already finite and equal to $8a_{0,2}^{\rm (GFF)}=8\Delta_\OO(\Delta_\OO+1)\zeta(2\Delta_\OO+2)$ without the need to regulate using the $\XX_T$ subtraction. The $\XX_T$ subtraction is the term in the last line which combines with the divergent sum/integral in the fourth line to produce a finite result.

More precisely, the add-back term on the fourth line
\begin{equation}
\label{b02}
\mathcal B_{0,2}(\mu)\coloneqq-\frac{2}{3}\NN_{\Delta_\OO} \frac{a_2^{(1)}}{\beta^4}\frac{\Gamma(\Delta_\OO)}{\Gamma(\Delta_\OO-4)}\frac{1}{\beta}\sum_{n\in \mathbb{Z}}\int_{\mathbb{R}^3} \frac{\dd^3 k}{(2\pi)^3} \left(1-e^{-\frac{\pp^2}{\mu^2}}\right)\;\GG_0(\pp)\frac{(3\omega_n^2-k^2)^2}{\pp^6}
\end{equation}
is equivalent to
\begin{equation}
    \mathcal{B}_{0,2}(\mu)=-\frac{\SS_{\Delta_\OO}}{24} \frac{1}{\beta}\sum_{n\in{\mathbb Z}} \int_{0}^{\infty} \dd k\; (1-e^{-\frac{p_n^2}{\mu^2}})\;p_n^{2\Delta_\OO-10}\; k^2(3\omega^2_n-k^2)^2 \; ,
\end{equation}
where we substituted the zero-temperature response \eqref{eq:G0} and evaluated the angular integral. This expression can be recast using Schwinger parametrization \eqref{eq:schwinger} into the form
\begin{equation}
\label{b02aa}
\BB_{0,2}(\mu) = -  
\frac{\SS_{\Delta_\OO}}{24\Gamma(5-\Delta_\OO)}\int_0^\infty \dd s \; s^{4-\Delta_\OO}\; \frac{1}{\beta} \sum_{n\in \mathbb Z} \left(I_n^{(2)}(s)- I_n^{(2)}(s+\mu^{-2}) \right)\; .
\end{equation}
Since
\begin{equation}
\label{sumJn}
\frac{1}{\beta}\sum_{n\in\mathbb Z} I_n^{(2)}(\alpha) = \frac{3}{4\alpha^4} + \frac{3}{2\alpha^4}\sum_{m=1}^\infty e^{-\frac{m^2\beta^2}{4\alpha}} - \frac{9\beta^2}{8\alpha^5}\sum_{m=1}^\infty m^2 e^{-\frac{m^2\beta^2}{4\alpha}} + \frac{9\beta^4}{64\alpha^6}\sum_{m=1}^{\infty}m^4e^{-\frac{m^2\beta^2}{4\alpha}}
\end{equation}
there is a divergence from the first term in \eqref{sumJn} when we substitute inside the $s$-integral in \eqref{b02aa}
\begin{equation}
\label{b02ab}
\BB_{0,2}(\mu) \supset - 
\frac{\SS_{\Delta_\OO}}{32\Gamma(5-\Delta_\OO)}\int_{\Lambda^{-2}}^\infty \dd s \; s^{-\Delta_\OO} = -\frac{\SS_{\Delta_\OO}}{32(\Delta_\OO-1)\Gamma(5-\Delta_\OO)} \Lambda^{2\Delta_{\OO}-2} 
~.
\end{equation}
where $\Lambda$ is a UV regulator.
On the other hand, the last line in \eqref{eq:a02_window_general} gives, 
\begin{equation}
\frac{\mathcal S_{\Delta_{\mathcal O}}}{96}
\int_0^\infty \dd p\,p^{2\Delta_{\mathcal O}-3}
\to
\frac{\mathcal S_{\Delta_{\mathcal O}}}{32\,\Gamma(5-\Delta_{\mathcal O})}
\int_{\Lambda^{-2}}^\infty \dd s\,s^{-\Delta_{\mathcal O}}
=\frac{\SS_{\Delta_\OO}}{32(\Delta_\OO-1)\Gamma(5-\Delta_\OO)} \Lambda^{2\Delta_{\OO}-2} 
~,
\end{equation}
where after the arrow we used the \emph{exact same} regulated Schwinger parameterization \eqref{eq:schwinger}. The remaining finite part can be calculated to be,
\begin{align}
     \mathcal B^{\rm(finite)}_{0,2}(\mu)
=\nonumber
&-
\frac{16a_2^{(1)}}{4^{\Delta_\OO} \beta^4}\,
\frac{\Gamma(2-\Delta_\OO)}{\Gamma(\Delta_\OO-4)}
\sum_{m=1}^{\infty}
\Bigg[
\frac{\Gamma(\Delta_\OO-1)(3\Delta_\OO^2-9\Delta_\OO+8)}
{4\Gamma(5-\Delta_\OO)}
\left(\frac{m^2\beta^2}{4}\right)^{1-\Delta_\OO}
\\[1mm]\nonumber
&\hspace{35mm}
-\frac{\Gamma(\Delta_\OO+1)}{160}
\left(\frac{m^2\beta^2}{4}\right)^2
\mu^{2\Delta_\OO+2}\,
{}_1F_1\!\left(
\Delta_\OO+1;6;-\frac{m^2\beta^2\mu^2}{4}
\right)
\\[1mm]\nonumber
&\hspace{35mm}
+\frac{\Gamma(\Delta_\OO)}{16}
\left(\frac{m^2\beta^2}{4}\right)
\mu^{2\Delta_\OO}\,
{}_1F_1\!\left(
\Delta_\OO;5;-\frac{m^2\beta^2\mu^2}{4}
\right)
\\[1mm]
&\hspace{35mm}
-\frac{\Gamma(\Delta_\OO-1)}{12}
\mu^{2\Delta_\OO-2}\,
{}_1F_1\!\left(
\Delta_\OO-1;4;-\frac{m^2\beta^2\mu^2}{4}
\right)
\Bigg].
 \end{align}
As a result, after collecting all the terms, the UV divergence cancels and we obtain the final expression

\begin{align}\label{eq:final_a02}
    a_{0,2} =  \mathcal R_0^{(0,2)}(\mu) + 2\sum_{n=1}^{\infty}\mathcal R_n^{(0,2)}(\mu)+ a_{0,2}^{\rm (GFF)} +  \beta^{2\Delta_\OO+2}\mathcal B^{\rm(finite)}_{0,2}(\mu)
\end{align}

where we defined,
 \begin{align}
    \mathcal  R_n^{(0,2)}(\mu) \coloneqq -\frac{\mathcal N_{\Delta_\OO}}{2 \pi^2 \beta} 
     \frac{\beta^{2\Delta_\OO+2}}{8}\int_{0}^{\infty} \dd k  \;k^2\left(\frac{3\omega_n^2-k^2}{3} \right)\times \cr
     \times\Bigg [\GG_T(n,k) -\GG_0(p_n) -  16\,a_2^{(1)}\frac{\Gamma(\Delta_\OO)}{\Gamma(\Delta_\OO-4)}&(1-e^{-p_n^2/\mu^2})\;\GG_0(p_n)\frac{3\omega_n^2-k^2}{\beta^4p_n^6} \Bigg] \;.\cr
 \end{align}

\begin{figure}[t]
  \centering
  \includegraphics[width=0.78\textwidth]{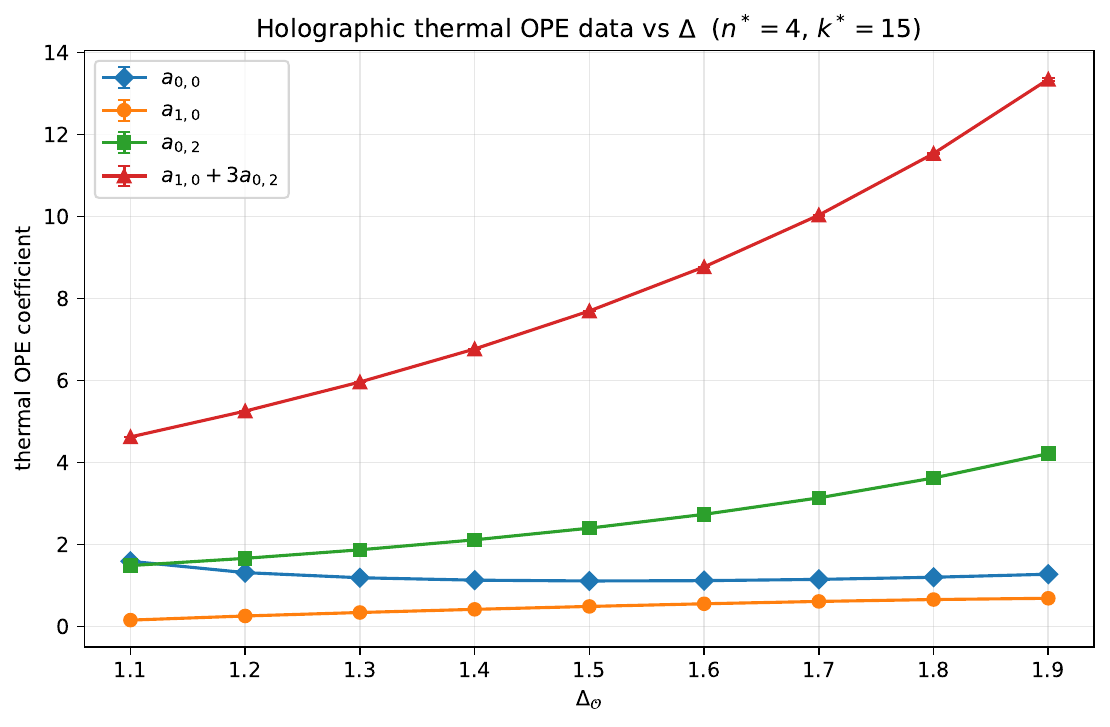}
  \caption{The holographic thermal double-twist data $a_{0,0},a_{1,0},a_{0,2}$, and the KMS
  combination $a_{1,0}+3a_{0,2}$ as functions of the external dimension $\Delta_\OO$ across the
  window $\Delta_\OO\in[1.1,1.9]$, at $(n^*,k^*)=(4,15)$. All four are smooth and analytic
  through $\Delta_\OO=\frac{3}{2}$. Error bars (often smaller than the markers) are estimated from the
  $k^*$ truncation and grow toward $\Delta_\OO\to2$, where $\GG_0\sim p^{-2|\nu|}$ falls off
  ever more slowly. The scalar coefficient $a_{0,0}$ attains a smooth minimum at
  $\Delta_\OO=\frac{3}{2}$.}
  \label{fig:all_vs_Delta}
\end{figure}

\subsection{Numerical implementation and results}
\label{sec:results}

We are finally in a position to assemble the finite results we have obtained in this section for $a_{0,0}$, $a_{1,0}$ and $a_{0,2}$. Each of the expressions \cref{eq:final_a00,eq:final_a10,eq:final_a02} contains closed analytic pieces (GFF and add-back contributions) and finite residuals involving the response $\GG_T(n,k)$ that need to be evaluated by numerical integration. To do so we deploy cutoffs $(n^*,k^*)$ to the Matsubara sum and the radial spatial momentum integral respectively, which introduces a truncation error.

\begin{figure}[t]
  \centering
  \includegraphics[width=0.78\textwidth]{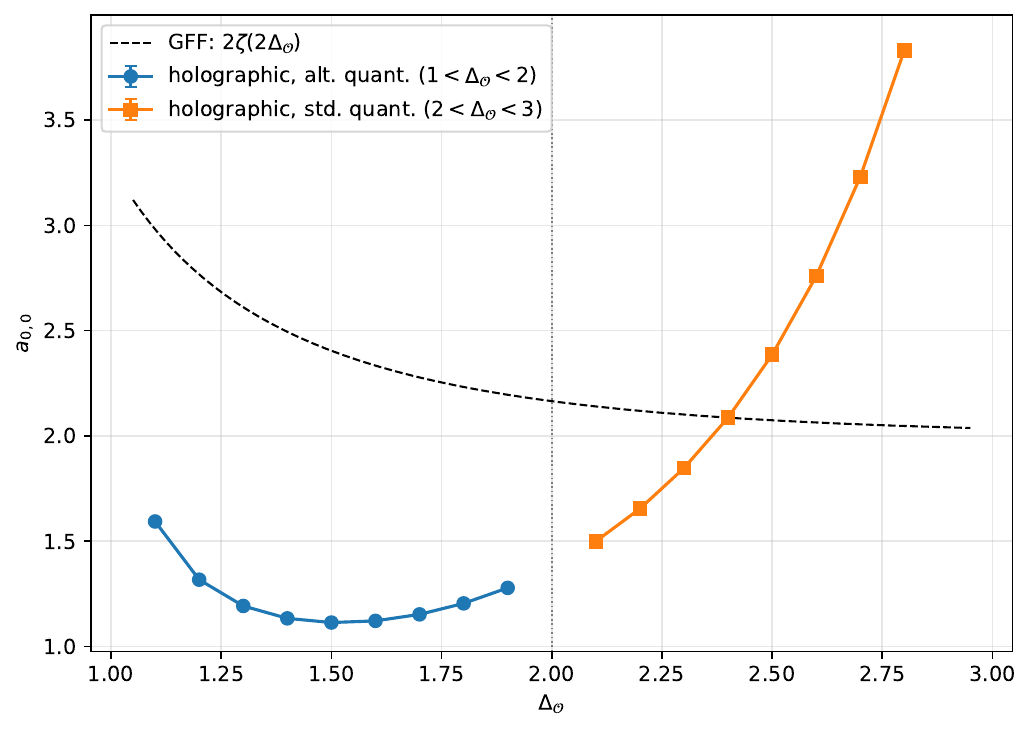}
  \caption{$a_{0,0}(\Delta_{\mathcal O})$ on both quantization branches of
the same bulk scalar ($m^2\ell^2=\Delta_{\mathcal O}(\Delta_{\mathcal O}-4)$). The dotted line marks the degenerate
point $\Delta_{\mathcal O}=2$ ($\nu=0$).}
  \label{fig:a00_vs_Delta}
\end{figure}

We first  carry out three diagnostics to confirm that the results are not an artifact of the finite-part prescription:
\begin{itemize}
  \item \emph{Scale independence.} Varying $\mu\in\{0.5,1,2\}$ shifts the analytic add-back by
  $\mathcal{O}(1)$ (e.g.\ $\mathcal{B}_{1,0}$ by a factor of six, $\mathcal{B}_{0,2}$ flipping
  sign), while each $a_{s,J}$ moves by $\lesssim 10^{-7}$. For each coefficient, the $\mu$-dependences of the residual and the add-back cancel against each other.
  
  \item \emph{Analyticity in $\Delta_\OO$.} Scanning $\Delta_\OO\in[1.1,1.9]$, the curves
  $a_{s,J}(\Delta_\OO)$ pass smoothly through $\Delta_\OO=\frac{3}{2}$ with neither a pole nor a
  slope discontinuity (see \cref{fig:all_vs_Delta} and \cref{fig:a00_vs_Delta}). A logarithm at $\Delta_\OO=\frac{3}{2}$ would have
  left a finite counterterm ambiguity in $a_{1,0},a_{0,2}$ individually. Its absence confirms
  that the prescription is scheme-independent.

\item \emph{Independence of the dressing function.} Replacing \eqref{eq:a00_window} by
$\tanh^2(p^2/\mu^2)$ or $1-e^{-p^4/\mu^4}$, which also satisfy \eqref{eq:window_props},
leaves $a_{0,0}$ unchanged for each $\mu\in\{0.5,1,2\}$, to the same accuracy as in the
scale-independence test. Profiles that approach
unity only as a power of $p$ are also admissible, but slow down the convergence of the
truncated sums.
  
\end{itemize}

Before we quote results evaluated with $\Delta_\OO=\frac{3}2$ ($\nu=\tfrac12$,
alternative quantization), $(n^*,k^*)=(15,20)$, $\beta=\pi$, $\mu=1$, we include a short discussion of our numerical sum/integration error. At fixed $k^*=20$ 
the $n^*>15$ tail is estimated at 
$\sim2\times10^{-4}$ for $a_{1,0}$, $a_{0,2}$ and $\sim 2\times10^{-7}$
for $a_{0,0}$. For $a_{0,0}$ 
the residual integrand over $k$ falls off as $k^{-7}$ (the $k^2$ moment), whereas
$a_{1,0},a_{0,2}$ involve integrands that decay only as $k^{-5}$ (the $k^4$ moment). The
dominant uncertainty in the numerical evaluation is therefore the $k^{*-4}$ tail of the latter two that contributes $\sim 7\times10^{-4}$ at
$k^*=20$. Pushing $k^*$ much further trades signal for numerical noise. The principled route to push below $10^{-3}$ would be to subtract analytically the
next multi-stress-tensor term ($T^2$) using the
input for the multi-stress-tensor data from the recursion of~\cite{Fitzpatrick:2019zqz}. This would shrink the
integrand falloff to $\sim k^{-9}$ and the $k^*$ tail to ${k^*}^{-8}$.

These statements are quantified for $a_{0,0}$ in \cref{tab:new_tab}. Each additional term of
\eqref{eq:GT_asym_general} lowers the contribution of the region outside the cutoffs by
several orders of magnitude. With the two terms used here this contribution is
$7\times10^{-7}$ at $k^*=20$, three times larger than the accumulated error of the numerical
solution of the radial equation, whereas at larger $k^*$ the latter dominates. This motivates
our choice of cutoffs.

\begin{table}[t]
\centering
\begin{tabular}{ccccc}
\toprule
 & \multicolumn{3}{c}{truncated tail} & numerical error\\
\cmidrule(lr){2-4}
$k^*$ & $\mathds{1}$ & $\mathds{1},\,T$ & $\mathds{1},\,T,\,T^2$ & \\
\midrule
$10$ & $5\times10^{-2}$ & $2\times10^{-5}$ & $3\times10^{-8}$ & $1\times10^{-7}$\\
$20$ & $2\times10^{-2}$ & $7\times10^{-7}$ & $<10^{-9}$ & $2\times10^{-7}$\\
$30$ & $7\times10^{-3}$ & $1\times10^{-7}$ & $<10^{-9}$ & $5\times10^{-7}$\\
\bottomrule
\end{tabular}
\caption{Contributions to $a_{0,0}$ at $\Delta_\OO=\tfrac32$ and $n^*=15$. The three middle
columns give the truncated tail when the terms of \eqref{eq:GT_asym_general} shown in the
header are subtracted. The last column gives the accumulated error of the numerical
solution of the radial equation.}
\label{tab:new_tab}
\end{table}

Putting everything together, we obtain:
\begin{equation}\label{eq:final_results}
  a_{0,0}=1.113079(1)\;,\qquad a_{1,0}=0.489(1)\;,\qquad a_{0,2}=2.399(1)\;.
\end{equation}
Here, the parenthetical is the uncertainty in the last quoted digit, dominated by the $k^*$ tail of the
residual, $\sim10^{-3}$ for $a_{1,0},a_{0,2}$ and $\sim10^{-6}$ for the faster-converging $a_{0,0}$.
These coefficients depart from their GFF values, the difference being the genuine black brane contribution captured by the residual
and add-back (for reference
$a^{\rm GFF}_{0,0}=2\zeta(2\Delta_\OO)=2\zeta(3)\simeq2.40)$. The mixed-spin combination singled out by the KMS (thermal periodicity)
condition~\cite{Niarchos:2025cdg},
\begin{equation}\label{eq:kms}
  a_{1,0}+3a_{0,2}=7.686(2)\;,
\end{equation}
carries a larger uncertainty than its constituents. In this combination the divergent fourth moment
cancels, but the finite $k^*$ tails of $a_{1,0}$ and $a_{0,2}$ do not. They leave a net truncation error $\sim1.4\times10^{-3}$, which
we quote conservatively as $2\times10^{-3}$. This result agrees very well with the zero-separation limit
evaluation of~\cite{Buric:2025anb,Buric:2025fye} ($\simeq7.686$). The scalar
value $a_{0,0}=1.113079(1)$ is consistent with the earlier holographic estimate
$\simeq1.1$ of~\cite{Parisini:2023nbd} and improves it to the 6th digit.

\section{Outlook} \label{sec:outlook}

In this work we have presented a method for extracting the thermal one-point
coefficients $a_{s,J}$ of the double-twist operators $[\OO\OO]_{s,J}$ that
appear in the scalar two-point function $\langle \OO(x)\OO(0)\rangle_\beta$
of a holographic CFT on $S^1_\beta \times \IR^{d-1}$ at large central charge.
Our main result is a formula that expresses each $a_{s,J}$ individually as a
regulated momentum-space integral of the thermal response function
$\GG_T(\omega_n,k)$. We derived this in two independent ways. In the first, we deformed the boundary theory by the corresponding double-trace operator
and read off the thermal one-point coefficient by differentiating the resulting
one-loop generating functional in the deformation coupling. In the second, we defined the same operator by
point-splitting. We acted on the thermal two-point function with the corresponding
differential operator, isolated the target block in the coincidence limit and
matched the result to the momentum-space integral of the thermal response
function. The only theory-dependent input
was $\GG_T$ itself, which follows from a radial reduction of the bulk Klein--Gordon
equation. For the case of a probe scalar in the planar AdS$_5$
black brane ($d=4$, leading order in $1/N$), it was obtained by numerically solving the resulting radial
ODE.

By choosing a representative non-integer dimension $\Delta_\OO = \tfrac{3}{2}$, we used this formula to 
determine the three lightest thermal coefficients $a_{0,0}$, $a_{1,0}$ and $a_{0,2}$.
The scalar datum $a_{0,0}$ was given to 7 significant figures and was compatible with the previous holographic calculation of \cite{Parisini:2023nbd}, while
$a_{1,0}$ and $a_{0,2}$ are, to our knowledge, the first determinations of these
individual coefficients. The significance of these numbers is
that they are spin-resolved data, inaccessible to the current bootstrap analysis. The KMS combination $a_{1,0}+3a_{0,2}$ agrees with the
zero-spatial-separation thermal bootstrap result of \cite{Buric:2025anb}.  

It would be useful to streamline the computation of higher double-twist thermal data. Reaching a heavier operator of dimension $2\Delta_\OO+2s+J$ requires subtracting the finite number of lighter multi-stress-tensor contributions below it (which can be obtained through \cite{Fitzpatrick:2019zqz,Karlsson:2019dbd,Karlsson:2022osn,Buric:2025fye}). At the same time, the regularization of the momentum integral \cref{eq:Wdef} becomes increasingly difficult: its UV divergence worsens with the dimension and
progressively more terms of the high-momentum asymptotic expansion of
$\GG_T$ must be subtracted to render it finite, leaving a small residual of an increasingly delicate cancellation. Since \cref{eq:GT_asym_general} involves factorially growing coefficients, the subtracted terms are individually large and the bulk response must be resolved to correspondingly higher accuracy. Thus, while the method is not limited in principle, the numerical cost grows steeply and the double precision used in our integration would need to be increased in practice to gain access beyond the leading double-twists.

More generally, one can explore various other related aspects of this work in holography. For example, our derivation of the master formula \eqref{eq:equiv_master} was essentially field-theoretic, with the bulk entering only through the input of the thermal response function $\GG_T$. It would be interesting to reformulate the derivation of the double-twist thermal one-point functions as a genuinely bulk computation. The one-loop generating
functional~\eqref{eq:Wlambda_O2} already contains a functional
determinant that one could evaluate directly from the bulk. This is what the
double-trace and mixed-boundary-condition framework of
\cite{Witten:2001ua,Gubser:2002zh,Gubser:2002vv,Mueck:2002gm,Papadimitriou:2007sj},
together with the functional-determinant technology developed for this purpose
in \cite{Hartman:2006dy,Diaz:2007an}, is designed to do. Applied to the AdS$_5$
black brane, such an approach would provide an alternative starting point for computing thermal double-twist data.

Another natural further direction would be to place the theory at finite spatial volume (on
$ S_\beta^1\times S^{d-1} $), replacing the planar black brane by the global
AdS$_5$--Schwarzschild black hole. Our construction should carry over once the
continuous spatial-momentum integrals are traded for discrete sums over the
hyperspherical harmonics on $S^{d-1}$ (the machinery of
Appendix~\ref{app:gegenbauer}), now coupled to the radial problem in the global
geometry. This would give access to genuinely finite-volume thermal data, making direct contact with recent studies of thermal CFT on $S_\beta^1\times S^{d-1}$~\cite{Fitzpatrick:2019zqz,Buric:2024kxo,David:2024pir,David:2025tqn,Barrat:2025twb,Buric:2026pes,Ammon:2025cdz}.

\section*{Acknowledgments}
We would like to thank Mitchell Woolley for many discussions and collaboration at the initial stages of this project, and Ioannis Papadimitriou for comments on the manuscript. The work of CP was partially supported by the Science and Technology Facilities Council (STFC) Consolidated Grant ST/X00063X/1 “Amplitudes, Strings \& Duality”. 

\begin{appendix}
\crefalias{section}{appendix}

\section{Holographic two-point function}
\label{app:holoren}
This appendix collects the holographic
renormalization that defines the renormalized action, the zero-temperature response
$\GG_0$, and its finite-temperature analog $\GG_T$. We work on the Euclidean planar AdS$_5$--Schwarzschild black brane~\eqref{eq:ads5bh} in $d=4$, with radial coordinate $z$, conformal boundary at $z=0$, horizon at $z=z_h$, AdS radius $\el$, and blackening factor $f(z)=1-z^4/z_h^4$. The zero-temperature limit is pure AdS$_5$, recovered as $z_h\to\infty$ ($f\to1$).

The bulk scalar in $d=4$ has the Euclidean action
\begin{equation}\label{eq:bulk_action}
  S_E=\frac12\int_\MM \dd^5x\,\sqrt{g}\,\bigl(\nabla_M\phi\,\nabla^M\phi+m^2\phi^2\bigr)\;,
\end{equation}
on $\MM=\{(x,z)\mid x\in\mathbb{R}^4,\ z\in(\epsilon,\infty)\}$ with the boundary
regulated at $z=\epsilon$. Integrating by parts,
\begin{equation}\label{eq:byparts}
  S_E=-\frac12\int_\MM \dd^5x\,\sqrt{g}\,\phi\,(\nabla_M\nabla^M-m^2)\phi
  +\frac12\int_{\partial\MM} \dd^4x\,\sqrt{\gamma}\,n^M\phi\,\nabla_M\phi\;,
\end{equation}
where $\gamma_{\mu\nu}=(\el^2/\epsilon^2)\delta_{\mu\nu}$ is the induced metric on the
cutoff surface $z=\epsilon$ and $n^M=-(\epsilon/\el)\,\delta^{Mz}$ is the outward unit
normal vector, pointing towards the boundary. The bulk term vanishes on shell and the action
reduces to the boundary term
\begin{equation}\label{eq:bdyaction}
  S_E^{\rm bdy}=-\frac12\int \dd^4 x\,\frac{\el^3}{z^3}\,\phi\,\partial_z\phi\,\Big|_{z=\epsilon}\;,
\end{equation}
which diverges as $\epsilon\to0$. Substituting the near-boundary
expansion~\eqref{eq:nearbd} and organizing by the source/response coefficients gives
\begin{align}\label{eq:divstructure}
  S_E^{\rm bdy}(\epsilon)=-\frac{1}{2}\int\frac{\dd^4p}{(2\pi)^4}\Bigl[
  &A(-p)A(p)\!\sum_{m,n\ge0}\!(2n+2-\nu)\,a_m a_n\,(p^2)^{m+n}\,\epsilon^{-2\nu+2m+2n}\nonumber\\
  &+A(-p)B(p)\!\sum_{m,n\ge0}\!2(m+n+2)\,a_m b_n\,(p^2)^{m+n}\,\epsilon^{2m+2n}\nonumber\\
  &+B(-p)B(p)\!\sum_{m,n\ge0}\!(2n+2+\nu)\,b_m b_n\,(p^2)^{m+n}\,\epsilon^{2\nu+2m+2n}\Bigr]\;.
\end{align}
The three contributions are bilinear in the source and response, schematically of the
form $A^2$, $AB$ and $B^2$. Here the two towers $a_m,b_m$ are the Frobenius coefficients of the near-boundary
branches $z^{4-\Delta_\OO}$ and $z^{\Delta_\OO}$ (or equivalently, the small-argument series
coefficients of the regular solution $\ell^{-3/2}(pz)^2K_\nu(pz)$ of \eqref{eq:physsoln}) given by\footnote{We assume $\Delta_\OO\notin\mathbb{N}$ (i.e.\ $\nu\notin\mathbb{Z}$)
as in the main text. For positive integer $\nu$, the Pochhammer symbol
$(1-\nu)_m$ vanishes for $m\geq \nu$, so the coefficient formula for
$a_m$ in~\eqref{eq:ambm} becomes singular; for $\nu=0$ the two near-boundary
branches degenerate. In either case the expansion~\eqref{eq:divstructure}
acquires logarithmic terms, signaling the holographic conformal anomaly.
These are removed by an additional logarithmic counterterm~\cite{deHaro:2000vlm}
and do not affect the renormalization structure used below.}
\begin{equation}\label{eq:ambm}
  a_m=\frac{1}{4^m\,m!\,(1-\nu)_m}\;,\qquad
  b_m=\frac{1}{4^m\,m!\,(1+\nu)_m}\;,
\end{equation}
where $(x)_m=\Gamma(x+m)/\Gamma(x)$ is the Pochhammer symbol.
The locally divergent contribution is the square of the leading
near-boundary branch, that is the coefficient of the smaller power $z^{\Delta_-}$,
$\Delta_-\equiv\min(\Delta_\OO,4-\Delta_\OO)$. The $AB$ cross term is
finite and non-local, while the square of the subleading branch vanishes as
$\epsilon\to0$. In our fixed-power labelling this leading coefficient is $A$ for
$\Delta_\OO>2$ and $B$ for $\Delta_\OO<2$ (so the divergent term is $A^2$ or
$B^2$ respectively), but the divergence structure is the same in both cases.

The divergences are removed by local
boundary counterterms
\begin{equation}\label{eq:counterterm}
  S_{\rm ct}^{(q)}=\int_{z=\epsilon}\dd^4 x\,\sqrt{\gamma}\,\phi\,(\Box_\gamma)^q\phi\;,
  \qquad q\in\mathbb{N}\;,
\end{equation}
where $\Box_\gamma$ is the Laplacian of the induced metric $\gamma_{\mu\nu}$.
Evaluated on the near-boundary expansion, each $S_{\rm ct}^{(q)}$ reproduces the
triple-sum structure of~\eqref{eq:divstructure} with every power of $\epsilon$ raised
by $2q$. In particular, its $AB$ cross-term scales as $\epsilon^{2q+2m+2n}$ and is
finite only for $q=m=n=0$. A single counterterm therefore overlaps the finite, non-local $AB$ piece, while the rest provide purely local subtractions. With coefficients $c_q$ fixed by minimal subtraction of the leading-branch
($z^{\Delta_-}$) divergence\footnote{This involves the same local counterterm in either
quantization.} the renormalized action
\begin{equation}\label{eq:Sren0}
  S_{\rm ren}=\lim_{\epsilon\to0}\Bigl(S_E^{\rm bdy}(\epsilon)
  +\sum_{q\ge0}c_q\,S_{\rm ct}^{(q)}(\epsilon)\Bigr),
\end{equation}
is the finite, non-local remainder
\begin{equation}\label{eq:Sren_result}
  S_{\rm ren}=-\nu\,\int\frac{\dd^4p}{(2\pi)^4}\,A(-p)\,B(p) = -\frac{1}{2}\, \int\frac{\dd^4p}{(2\pi)^4}\, A(-p) \GG_0(p) A(p)\;,
  \quad \nu = |\Delta_\OO-2|\;.
\end{equation}
The AdS/CFT dictionary~\cite{Gubser:1998bc,Witten:1998qj}, $Z[A]=e^{-S_{\rm ren}[A]}$, and \eqref{eq:G0} then gives the renormalized conjugate momentum\footnote{In general the renormalized momentum takes the form
$\Pi=(\Delta_+-\Delta_-) B+C[A]$, with $\Delta_\pm=2\pm\nu$ and $C$ a local,
generically non-linear, functional of the source~\cite{Papadimitriou:2007sj}. As shown
there, $C$ vanishes identically for the boundary conditions relevant to multi-trace
deformations, which is precisely the case of interest here, so that
$\Pi=2\nu\, B$ exactly. For non-integer $\Delta_\OO$ this is also evident
from~\eqref{eq:divstructure}: a finite local ($A^2$) term would require $m+n=\nu$.}
\begin{align}
  -\frac{\delta S_{\rm ren}}{\delta A(-p)}
 =   \langle \OO(p)\rangle  = \GG_0(p) A(p)\;.
\end{align}
The momentum-space
two-point function follows
\begin{equation}\label{eq:2pt}
  \langle\OO(q)\,\OO(p)\rangle
  =\frac{\delta \langle \OO(p)\rangle}{\delta A(-q)}\Big|_{A=0}
  =(2\pi)^4\,\GG_0(p)\,\delta^{(4)}(p+q)\;,
\end{equation}
and, in position space,
\begin{equation}\label{eq:2ptposition}
  \langle\OO(x)\,\OO(0)\rangle
  =\frac{2\nu(\Delta_\OO-2)(\Delta_\OO-1)}{\pi^2}\,\frac{1}{|x|^{2\Delta_\OO}}\;.
\end{equation}
This is written in the bulk-canonical normalization inherited from~\eqref{eq:bulk_action}.
Rescaling $\OO$ by the constant appearing in~\eqref{eq:2ptposition} recovers the
unit-normalized convention~\eqref{eq:2ptpositionmain} used in the main text, with
$\NN_{\Delta_\OO}$ as in~\eqref{eq:Ndef}.
This derivation is in standard quantization. For the alternative quantization of the main text
($\Delta_\OO<2$) source and response are exchanged, the response being the reciprocal~\eqref{eq:G0},
and the renormalized correlator is obtained from the momentum conjugate to the new source. The
overall constant differs from the standard-quantization result, but it is again absorbed by the
unit normalization~\eqref{eq:2ptpositionmain}, which is what the main text uses.

At finite temperature the conformal boundary is unchanged, and the induced metric
on the cutoff surface $z=\epsilon$ differs from its zero-temperature form only
through the blackening factor $f(\epsilon)=1-\epsilon^4/z_h^4=1+\mathcal{O}(\epsilon^4)$.\footnote{Explicitly, $\gamma_{\mu\nu}=\frac{\el^2}{\epsilon^2}\bigl(f(\epsilon)\,
\delta_{\mu0}\delta_{\nu0}+\delta_{\mu i}\delta_{\nu i}\bigr)$, so that
$\sqrt{\gamma}=\frac{\el^4}{\epsilon^4}\sqrt{f(\epsilon)}$ and the outward unit
normal is $n^M=-\frac{z}{\el}\sqrt{f(\epsilon)}\,\delta^{zM}$. All reduce to their
zero-temperature values as $\epsilon\to0$.}
Since the divergences are controlled by the leading near-boundary behavior, they
are insensitive to $f$: the thermal corrections first modify the Frobenius
coefficients at order $z^4$ (see \cref{app:frobenius}) and contribute only finite,
non-local terms. The same local counterterms therefore renormalize the thermal
action and the above steps go through verbatim under
$\int \dd^4p/(2\pi)^4\to\beta^{-1}\sum_n\int \dd^3k/(2\pi)^3$ and $\GG_0\to\GG_T$. This yields
the renormalized action~\eqref{eq:SrenBH} and the bulk thermal two-point function~\eqref{eq:gGT}.

\section{Useful identities}
\label{app:identities}
This appendix contains a collection of technical expressions used in the main text: the Frobenius
coefficients of the bulk radial problem of \cref{sec:radialT}, the high-momentum
asymptotics of the thermal blocks (which control the subtractions), the minimal
number of subtractions, and the Gegenbauer/hyperspherical projection that isolates a
single spin contribution. Throughout, $\nu=|\Delta_\OO-2|$ is the bulk Bessel order of
\cref{sec:setup} and $\hat\nu=(d-2)/2$ the Gegenbauer index, with $\hat\nu=1$ in $d=4$.

\subsection{Bulk radial Frobenius coefficients}
\label{app:frobenius}
Here we list the recursion relations for the Frobenius coefficients of the bulk
radial equation~\eqref{eq:radialBH}, used in the near-boundary matching of
\cref{sec:radialT,sec:GTnumeric}.

\begin{table}[t]
\centering
\renewcommand{\arraystretch}{1.8}
\begin{tabular}{c|ccc}
\hline
 & $m=0$ & $m=2$ & $m=4$ \\
\hline
$a_m$ & $1$ & $\dfrac{\omega_n^2+k^2}{4(1-\nu)}$ &
  $\dfrac{(\omega_n^2+k^2)^2}{32(1-\nu)(2-\nu)}+\dfrac{2-\nu}{8z_h^4}$ \\
$b_m$ & $1$ & $\dfrac{\omega_n^2+k^2}{4(1+\nu)}$ &
  $\dfrac{(\omega_n^2+k^2)^2}{32(1+\nu)(2+\nu)}+\dfrac{\nu+2}{8z_h^4}$ \\
\hline
\end{tabular}
\caption{Lowest near-boundary Frobenius coefficients of the two branches
in~\eqref{eq:bdr_frob}. At large $n$ or $k$ the Bessel part dominates and the solution
reduces to its zero-temperature form. The finite-temperature corrections first enter at
$\mathcal{O}(z^4)$, through the $z_h^{-4}$ terms.}
\label{tab:frob}
\end{table}

\paragraph{Near-boundary coefficients.}
Substituting the near-boundary series~\eqref{eq:bdr_frob} into~\eqref{eq:radialBH}
yields two decoupled recursions for the branches $z^{2\mp\nu}$,
\begin{align}
  m(m-2\nu)\,a_m
  &-(\omega_n^2+k^2)\,a_{m-2}
  -\frac{2(m-\nu-2)(m-\nu-4)+(4-\nu^2)}{z_h^4}\,a_{m-4}\nonumber\\
  &+\frac{k^2}{z_h^4}\,a_{m-6}
  +\frac{(m-\nu-6)^2}{z_h^8}\,a_{m-8}=0\;,\label{eq:bdr_frob_a}\\
  m(m+2\nu)\,b_m
  &-(\omega_n^2+k^2)\,b_{m-2}
  -\frac{2(m+\nu-2)(m+\nu-4)+(4-\nu^2)}{z_h^4}\,b_{m-4}\nonumber\\
  &+\frac{k^2}{z_h^4}\,b_{m-6}
  +\frac{(m+\nu-6)^2}{z_h^8}\,b_{m-8}=0\;,\label{eq:bdr_frob_b}
\end{align}
with $a_0=b_0=1$ and $a_{m<0}=b_{m<0}=0$. Setting $m=1$ shows that all odd-$m$
coefficients vanish, and in the limit $z_h\to\infty$ the recursions reduce to those of
the modified Bessel functions $I_{\pm\nu}(\pp z)$, with $\pp=\sqrt{\omega_n^2+k^2}$. The
blackening factor enters only through the $z_h^{-4}$ and $z_h^{-8}$ terms, so the
leading thermal corrections first appear at $m=4$. The lowest coefficients are listed
in \cref{tab:frob}.

\paragraph{Near-horizon coefficients.}
In the dimensionless coordinate $u=z/z_h$, substituting the near-horizon
series~\eqref{eq:hor_frob} into~\eqref{eq:radialBH_u} yields the eighth-order recurrence
\begin{equation}\label{eq:hor_frob_rec}
  16\,m(m+|n|)\,h_m+\sum_{j=1}^{8}Q_j(m)\,h_{m-j}=0\;,\qquad h_{m<0}=0\;,
\end{equation}
with coefficients
\begin{align}
  Q_1(m)&=-4\bigl(20m^2+20m|n|-46m+3n^2-23|n|+k^2z_h^2+\nu^2+22\bigr)\;,\cr
  Q_2(m)&=180m^2+180m|n|-796m+41n^2-398|n|+14k^2z_h^2+6\nu^2+848\;,\cr
  Q_3(m)&=-4\bigl(60m^2+60m|n|-386m+15n^2-193|n|+5k^2z_h^2+\nu^2+614\bigr)\;,\cr
  Q_4(m)&=208m^2+208m|n|-1744m+52n^2-872|n|+15k^2z_h^2+\nu^2+3644\;,\cr
  Q_5(m)&=-6\bigl(20m^2+20m|n|-206m+5n^2-103|n|+k^2z_h^2+530\bigr)\;,\cr
  Q_6(m)&=\tfrac14\bigl(180m^2+180m|n|-2196m+45n^2-1098|n|+4k^2z_h^2+6696\bigr)\;,\cr
  Q_7(m)&=-\tfrac12\bigl(20m^2+20m|n|-282m+5n^2-141|n|+994\bigr)\;,\cr
  Q_8(m)&=\tfrac14(2m+|n|-16)^2\;.
\end{align}
These fix the $h_m$ recursively in terms of $h_0$, the overall normalization of the
regular near-horizon solution~\eqref{eq:hor_frob}.

\subsection{High-momentum asymptotics of the thermal OPE}
\label{app:highmom}
The thermal block expansion~\eqref{eq:thermal_OPE} of \cref{sec:deformations} is a small-$|x|$
expansion with radius of convergence $\beta$. Its Fourier transform does not reproduce
the full momentum-space correlator but, being a short-distance expansion, controls the large-momentum asymptotics, where the integrals are dominated by the
OPE region; we follow~\cite{Manenti:2019wxs}. The momentum-space transform of a single
block,
\begin{equation}\label{eq:largemomblock}
  G_{\Delta,J}^{(\Delta_\OO,\hat\nu)}(\omega_n,k)
  =\beta^{-\Delta}\int_{\mathbb{R}^{d-1}}\!d^{d-1}x\int_0^\beta\!d\tau\,
  e^{-i\omega_n\tau-i\bk\cdot\bx}\,C_J^{(\hat\nu)}\!\Bigl(\tfrac{\tau}{|x|}\Bigr)\,
  |x|^{-2\Delta_\OO+\Delta}\;,
\end{equation}
has the large-momentum asymptotics
\begin{align}\label{eq:largemomblock_eval}
  G_{\Delta,J}^{(\Delta_\OO,\hat\nu)}(\omega_n,k)
  ={}&\beta^{-\Delta}\sum_{j=0}^{\lfloor J/2\rfloor}(-1)^j
  \frac{\Gamma(J-j+\hat\nu)}{\Gamma(\hat\nu)\,j!\,(J-2j)!}\,2^{J-2j}\nonumber\\
  &\times\frac{2^{q_j}\pi^{\hat\nu+1}\Gamma(b_j)}{\Gamma(c_j)}\,
  \frac{{}_2F_1\!\bigl(q_j,1-q_j;1-b_j;\tfrac{k+i\omega_n}{2k}\bigr)}
  {k^{q_j}(k+i\omega_n)^{b_j}}+\mathcal{O}(e^{-\beta k})\;,
\end{align}
with
\begin{equation}
  q_j=\hat\nu+1+\tfrac{\Delta}{2}-\Delta_\OO-\tfrac J2+j,\quad
  b_j=\hat\nu+1+\tfrac{\Delta}{2}-\Delta_\OO+\tfrac J2-j,\quad
  c_j=\Delta_\OO+\tfrac J2-j-\tfrac{\Delta}{2}\;.
\end{equation}
A key property is that the large-momentum asymptotics vanish for every
double-twist operator. In particular, their explicit contribution in~\eqref{eq:largemomblock_eval} is zero, so that
\begin{equation}\label{eq:dt_vanish}
  G_{2\Delta_\OO+2s+J,\,J}^{(\Delta_\OO,\hat\nu)}=\mathcal{O}(e^{-\beta k})\;,
  \qquad s,J\in\mathbb{Z}_{\ge0}\;,
\end{equation}
for all $\Delta_\OO$ and $d$. Hence, for holographic CFTs the large-momentum sector is dominated by the
stress-tensor sector of the OPE. For a multi-stress-tensor operator~\cite{Karlsson:2019dbd,Karlsson:2022osn} with
$\Delta=4m$ in $d=4$,
\begin{align}\label{eq:EMblock}
  G_{4m,J}^{(\Delta_\OO,1)}(\omega_n,k)
  ={}& \frac{1}{2\nu}\beta^{-4m}\pi^2\,2^{4m}(-1)^{J/2}\,
  \frac{\Gamma\!\bigl(2m+\tfrac J2+2-\Delta_\OO\bigr)}{\Gamma(2-\Delta_\OO)}\,
  \frac{\Gamma(\Delta_\OO-2)}{\Gamma\!\bigl(\Delta_\OO-2m+\tfrac J2\bigr)}\nonumber\\
  &\times \pp^{-4m}\,C_J^{(1)}\!\Bigl(\tfrac{\omega_n}{\pp}\Bigr)\,\GG_0(\pp)
  +\mathcal{O}(e^{-\beta k})\;,
\end{align}
where we substituted the zero-temperature response~\eqref{eq:G0}. In particular,
for the stress tensor itself ($\Delta=4$, $J=2$),
\begin{align}\label{eq:Tblock}
  G_{4,2}^{(\Delta_\OO,1)}(\omega_n,k)
  &=\frac{16\pi^2}{2\nu}(\Delta_\OO-3)(\Delta_\OO-4)\,\frac{3\omega_n^2-k^2}{\beta^4 \pp^6}\,\GG_0(\pp)+\mathcal{O}(e^{-\beta k})\cr
  &=16\,\NN_{\Delta_\OO}\,\frac{\Gamma(\Delta_\OO)}{\Gamma(\Delta_\OO-4)}\,
  \frac{3\omega_n^2-k^2}{\beta^4 \pp^6}\,\GG_0(\pp)+\mathcal{O}(e^{-\beta k})\;,
\end{align}
with $\NN_{\Delta_\OO}$ as in~\eqref{eq:Ndef}. This is the asymptotic
subtracted to regularize $a_{1,0}$ and $a_{0,2}$ in \cref{sec:regularization}. In general, the holographic response function used in the main text admits the large-momentum  expansion
\begin{equation}\label{eq:GT_asym_general}
\mathcal{G}_T(\omega_n,k)
=
\sum_{m=0}^{\infty}\sum_{\substack{J=0\\ J\ \mathrm{even}}}^{2m}a_J^{(m)}
(-1)^{J/2}\left(\frac{2}{\beta \pp}\right)^{4m}\frac{\Gamma(\Delta_\OO)\Gamma\!\left(2+2m-\Delta_\OO+\frac{J}{2}\right)
}{\Gamma(2-\Delta_\OO)\Gamma\!\left(\Delta_\OO-2m+\frac{J}{2}\right)}C_J^{(1)}\!\left(\frac{\omega_n}{\pp}\right)\mathcal{G}_0(\pp)
\;.
\end{equation}
This  expansion is asymptotic rather than convergent, with  the order-$m$ coefficient growing like $[(2m)!]^2$. This does not affect  the subtraction scheme
of \cref{sec:window} since only a finite number of terms is removed. At momentum $p_n$ successive terms decrease up to the order $m\simeq\beta p_n/(4\sqrt2)$, so
that for the cutoffs used in this paper additional terms improve the convergence up to $m\simeq10$.

\subsection{Gegenbauer polynomials, hyperspherical harmonics, and plane waves}
\label{app:gegenbauer}
The Gegenbauer polynomials are orthogonal on $S^{d-1}$,
\begin{equation}\label{eq:gegen_ortho}
  \int_{S^{d-1}}d\Omega_{d-1}\,C^{(\hat\nu)}_\ell(\eta)\,C^{(\hat\nu)}_J(\eta)=N_J\,\delta_{\ell,J}\;,
  \qquad N_J=\frac{4^{1-\hat\nu}\pi^{\hat\nu+\frac32}\,\Gamma(J+2\hat\nu)}
  {J!\,(J+\hat\nu)\,\Gamma(\hat\nu)^2\,\Gamma(\hat\nu+\tfrac12)}\;.
\end{equation}
Another useful fact, used in \cref{sec:gendeform} to reduce the tensor
$\mathscr{P}_J$ to its angular average, is the zonal harmonic
identity. The tensor
$\mathscr{P}_J^{\mu_1\cdots\mu_J}(\pp)=\pp^{\mu_1}\cdots \pp^{\mu_J}-(\text{traces})$ is harmonic
and homogeneous of degree $J$ in $\pp$, so as a function of the direction $\hat p_n$ it is a
degree-$J$ harmonic on $S^{d-1}$ that depends only on $\hat p_n\cdot\hat a$. The unique such
zonal harmonic is the Gegenbauer polynomial $C_J^{(\hat\nu)}(\hat p_n\cdot\hat a)$ with
$\hat\nu=(d-2)/2$, so the two sides of~\eqref{eq:harmonic_id} agree up to a constant. That
constant is fixed by the coefficient of $(\hat p_n\cdot\hat a)^J$. On the left this top term
is $(\pp\cdot\hat a)^J=\pp^J(\hat p_n\cdot\hat a)^J$ with coefficient unity, while the leading
coefficient of $C_J^{(\hat\nu)}$ is $2^J(\hat\nu)_J/J!$, and the ratio reproduces the factor
$J!/[2^J(\hat\nu)_J]$.
In the main text we evaluate angular integrals of the form
\begin{equation}\label{eq:IJdef}
  I_J(r)=\int_{S^{d-1}}d\Omega_{d-1}\,C^{(\hat\nu)}_J(\eta)\,\frac1\beta\sum_{n\in\mathbb{Z}}
  \int_{\mathbb{R}^{d-1}}\frac{d^{d-1}k}{(2\pi)^{d-1}}\,e^{i\pp\cdot x}\,f(n,k)\;.
\end{equation}
Expanding the plane wave with the $d$-dimensional Rayleigh formula,
\begin{equation}\label{eq:rayleigh}
  e^{i\pp\cdot x}=2^{\hat\nu}\Gamma(\hat\nu)\sum_{\ell=0}^\infty i^\ell(\ell+\hat\nu)\,
  \frac{J_{\ell+\hat\nu}(\pp r)}{(\pp r)^{\hat\nu}}\,C_\ell^{(\hat\nu)}\!\Bigl(\tfrac{\pp\cdot x}{\pp r}\Bigr)\;,
\end{equation}
where $J_\alpha$ is the Bessel function of the first kind, gives
\begin{equation}\label{eq:IJexpanded}
  I_J(r)=2^{\hat\nu}\Gamma(\hat\nu)\,\frac1\beta\sum_{n\in\mathbb{Z}}
  \int_{\mathbb{R}^{d-1}}\frac{d^{d-1}k}{(2\pi)^{d-1}}\,f(n,k)
  \sum_{\ell=0}^\infty i^\ell(\ell+\hat\nu)\frac{J_{\ell+\hat\nu}(\pp r)}{(\pp r)^{\hat\nu}}\,X_{J,\ell}(n,k)\;,
\end{equation}
where
\begin{equation}\label{eq:XJl}
  X_{J,\ell}(n,k)=\int_{S^{d-1}}d\Omega_{d-1}\,
  C^{(\hat\nu)}_J(\hat\tau\cdot\hat x)\,C_\ell^{(\hat\nu)}(\hat p_n \cdot\hat x)\;,
  \qquad \hat\tau^\mu\coloneqq\delta^{\mu,0}\;.
\end{equation}
Using the hyperspherical harmonic addition theorem,
\begin{equation}\label{eq:addition}
  \sum_{m=1}^{Q_J^{(\hat\nu)}}Y_{J,m}(\hat a)\,Y_{J,m}^*(\hat x)
  =\frac{J+\hat\nu}{\hat\nu\,\Omega_{d-1}}\,C_J^{(\hat\nu)}(\hat a\cdot\hat x)\;,
  \qquad Q_J^{(\hat\nu)}=\frac{2(J+\hat\nu)(J+2\hat\nu-1)!}{J!\,(2\hat\nu)!}\;,
\end{equation}
with $\Omega_{d-1}=2\pi^{d/2}/\Gamma(d/2)$ and $Y_{J,m}$ the orthonormal
eigenfunctions of $-\nabla^2_{S^{d-1}}$ with eigenvalue $J(J+d-2)$, the angular
integral collapses to
\begin{equation}\label{eq:XJl_eval}
  X_{J,\ell}(n,k)=\frac{\hat\nu\,\Omega_{d-1}}{J+\hat\nu}\,
  C_J^{(\hat\nu)}\!\Bigl(\tfrac{\omega_n}{\pp}\Bigr)\delta_{\ell,J}
  =\frac{2\pi^{\hat\nu+1}}{(J+\hat\nu)\Gamma(\hat\nu)}\,
  C_J^{(\hat\nu)}\!\Bigl(\tfrac{\omega_n}{\pp}\Bigr)\delta_{\ell,J}\;,
\end{equation}
which for $\hat p_n=\hat\tau$ reduces to~\eqref{eq:gegen_ortho}. Substituting back and
assuming $f$ carries no angular dependence in the spatial momentum,\footnote{If $f$ carries Lorentz indices, the index structure enters through the
tensor $\mathscr{P}_J(\pp)$, which the spatial $O(3)$ average reduces to the invariant
$\mathcal{Q}_J$ times a scalar, see~\eqref{eq:PJaverage}.} 
the
projection reduces to a single radial integral
\begin{equation}\label{eq:gegen_fourier_proj_final}
  I_J(r)=\frac{2^{1-2\hat\nu}\sqrt\pi}{\Gamma(\hat\nu+\tfrac12)}\,\frac{(-1)^{J/2}}{\beta}
  \sum_{n\in\mathbb{Z}}\int_0^\infty \dd k\,k^{2\hat\nu}\,f(n,k)\,
  \frac{J_{J+\hat\nu}(\pp r)}{(\pp r/2)^{\hat\nu}}\,C_J^{(\hat\nu)}\!\Bigl(\tfrac{\omega_n}{\pp}\Bigr)\;.
\end{equation}
The Bessel fraction admits the series representation
\begin{equation}\label{eq:besselfraction}
  \frac{J_{J+\hat\nu}(\pp r)}{(\pp r/2)^{\hat\nu}}
  =\sum_{m=0}^\infty\frac{(-1)^m}{m!\,\Gamma(m+J+\hat\nu+1)}\Bigl(\frac{\pp r}{2}\Bigr)^{2m+J}\;,
  \qquad
  \lim_{r\to0}\frac{J_{J+\hat\nu}(\pp r)}{(\pp r/2)^{\hat\nu}}=\frac{\delta_{0,J}}{\Gamma(\hat\nu+1)}\;,
\end{equation}
so the coincidence limit $r \to 0$ isolates the spin-zero component,
\begin{equation}\label{eq:IJ0}
  I_J(0)=\delta_{J,0}\,\frac{2^{1-2\hat\nu}\sqrt\pi}{\Gamma(\hat\nu+\tfrac12)\Gamma(\hat\nu+1)}\,
  \frac1\beta\sum_{n\in\mathbb{Z}}\int_0^\infty \dd k\,k^{2\hat\nu}\,f(n,k)\;.
\end{equation}

\end{appendix}

\bibliographystyle{JHEP}
\bibliography{finiteT}

\end{document}